\documentclass[11pt]{article}

\usepackage[final]{acl}

\makeatletter
\ifdefined\@LN@column
  \RemoveFromHook{build/column/before}[lineno]
  \AddToHook{build/column/after}{%
    \setbox\@outputbox\vbox{%
      \boxmaxdepth\@maxdepth
      \protected@write\@auxout{}{%
        \string\def\string\@LN@column{\if@firstcolumn1\else2\fi}}%
      \box\@outputbox}}
\fi
\makeatother

\usepackage{times}
\usepackage{latexsym}

\usepackage[T1]{fontenc}

\usepackage[utf8]{inputenc}
\usepackage{svg}
\usepackage{microtype}

\usepackage{inconsolata}

\usepackage{graphicx}
\usepackage{amsmath}
\usepackage{amssymb}
\usepackage{booktabs}
\usepackage{multirow}
\usepackage{fontawesome5}
\usepackage{colortbl}
\usepackage{CJKutf8}
\newcommand{\zh}[1]{\begin{CJK}{UTF8}{gbsn}#1\end{CJK}}

\usepackage[most]{tcolorbox}
\usepackage{listings}

\usepackage{tcolorbox}

\newtcolorbox[auto counter]{findingbox}{
    enhanced,
    colback=gray!6,
    colframe=teal!65!black,
    boxrule=0.55pt,
    arc=1.8mm,
    left=6pt,
    right=6pt,
    top=4pt,
    bottom=4pt,
    before skip=5pt,
    after skip=5pt,
    fontupper=\normalfont,
    before upper={\textbf{\textit{Finding~\thetcbcounter.}}\ }
}

\newcommand{\dataset}{VoxCanvas-41M}
\newcommand{\model}{VoxCanvas-Edit}

\title{\dataset{}: A Large-Scale Bilingual Dataset for Instruction-Based Speech Editing and Its Learning Dynamics}

\author{
  \textbf{Zhikang Niu}\textsuperscript{1,2}
  \thanks{Equal contribution.} \quad
  \textbf{Wenming Tu}\textsuperscript{1,3}\footnotemark[1] \quad
  \textbf{Yalin Li}\textsuperscript{1,2} \quad
  \textbf{Yanru Huo}\textsuperscript{1,3} \quad
  \textbf{Junxi Liu}\textsuperscript{1,2} \quad
  \textbf{Tianrui Wang}\textsuperscript{4} \\
  \textbf{Ruiqi Yan}\textsuperscript{1} \quad
  \textbf{Junxuan Liang}\textsuperscript{1,2} \quad
  \textbf{Zilong Zheng}\textsuperscript{3} \quad
  \textbf{Ziyang Ma}\textsuperscript{1,2} \quad
  \textbf{Xie Chen}\textsuperscript{1,2}
  \thanks{Corresponding author.} \\
  \textsuperscript{1}X-LANCE Lab, Shanghai Jiao Tong University, China \\
  \textsuperscript{2}Shanghai Innovation Institute, China \\
  \textsuperscript{3}Beijing Institute for General Artificial Intelligence, China \\
  \textsuperscript{4}Tianjin University, China \\
  \href{mailto:zhikangniu@sjtu.edu.cn}{zhikangniu@sjtu.edu.cn};
  \href{mailto:chenxie95@sjtu.edu.cn}{chenxie95@sjtu.edu.cn}
}

\begin{document}
\maketitle
\begin{abstract}
Recent advances in speech generation have spurred the development of speech editing techniques. However, achieving high-quality speech editing remains challenging due to two key limitations. First, the major hurdle is the lack of large-scale, high-quality speech editing datasets. Constructing such datasets is costly, while open-source, scalable data pipelines remain limited. Second, the learning dynamics of speech editing models remain poorly understood. In particular, how models acquire editing capabilities and the relationship between generation and editing remain unclear. To address these challenges, we introduce \textbf{\dataset{}}, a large-scale bilingual dataset comprising approximately 41 million speech editing pairs. \dataset{} covers two major categories: content editing, including single-span, multi-span, and location-reference editing, and paralinguistic editing, covering the insertion and deletion of 39 classes of nonverbal events such as laughter, sighs, and coughs. Building on \dataset{}, we train \model{}, an instruction-based speech editing model, demonstrating the effectiveness of our dataset in improving editing performance. Furthermore, we systematically investigate the learning dynamics of speech editing, examining how models learn different editing operations and how data scale, task interactions, and training strategies affect editing performance. The dataset\footnote{Dataset: \href{https://huggingface.co/datasets/zkniu/VoxCanvas}{huggingface.co/datasets/zkniu/VoxCanvas}},
pipeline\footnote{Pipeline: \href{https://github.com/ZhikangNiu/VoxCanvas}{github.com/ZhikangNiu/VoxCanvas}},
and demos\footnote{Demo: \href{https://zhikangniu.github.io/voxcanvas/}{zhikangniu.github.io/voxcanvas/}}
are publicly available.
\end{abstract}

\begin{figure*}[t]
  \centering
  \includegraphics[width=\textwidth]{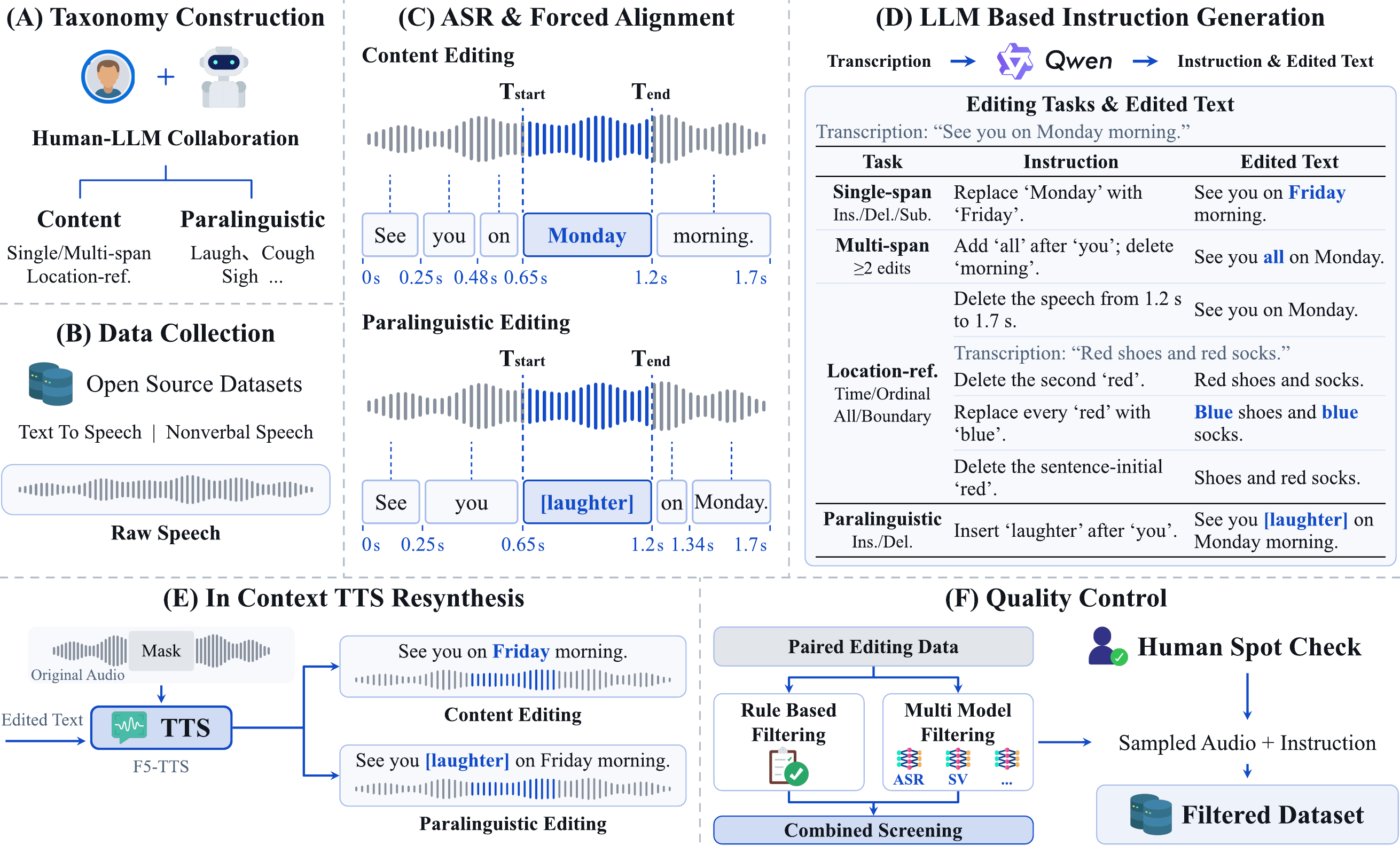}
  \caption{Overview of the speech editing data construction pipeline. ASR: automatic speech recognition; LLM: large language model; TTS: text-to-speech.}
  \label{fig:speech-editing-pipeline}
\end{figure*}

\section{Introduction}
\label{sec:intro}

Speech editing aims to modify existing speech while preserving the speaker identity, prosody, and acoustic environment of its unedited regions \citep{jin2017voco, peng2024voicecraft,yan2025stepaudioeditx,ma2026auk}.
It supports applications such as the post-production of podcasts and audiobooks, video dubbing, and the correction of errors in synthesized speech, all without regenerating the entire audio.
Existing speech editing methods can be broadly divided into two categories: infilling-based methods \citep{peng2024voicecraft, wang2025ssrspeech, zheng2025voicecraftx} and instruction-based methods \citep{huang2024instructspeech, yan2025stepaudioeditx, yan2025muniaudio,ma2026auk,wang2026dotsttsedit}.

\textbf{Representative infilling-based methods} learn editing as in-context masked infilling and achieve strong editing performance \citep{le2023voicebox, peng2024voicecraft, wang2025ssrspeech, zheng2025voicecraftx,chen2025f5tts,han2026fish}.
During training, they randomly mask spans of training speech and predict the masked spans from the transcript and the unmasked audio context.
Although this paradigm needs no paired editing data for training, it requires the edited transcript and the precise span of speech to regenerate at inference.
To obtain this span, these systems rely on external automatic speech recognition (ASR) and forced alignment tools such as Whisper \citep{radford2023whisper} and MFA \citep{mcauliffe2017mfa}.
\textbf{Instruction-based methods} learn to edit original speech from natural-language instructions without an edited transcript and precise edit span. This paradigm not only removes the need for external alignment tools but also enables flexible combinations of multiple types of edits.
However, these methods rely on large-scale, high-quality paired editing data, which is often not publicly available.
Step-Audio-EditX \citep{yan2025stepaudioeditx}, Ming-UniAudio \citep{yan2025muniaudio}, dots.tts.edit \citep{wang2026dotsttsedit}, and AuK \citep{ma2026auk} all train with this paradigm on in-house editing data. Among publicly available datasets, GigaEdit-L \citep{chen2026cosyedit} provides approximately 2.1M English editing pairs (2.2k hours), constructed through rule-based cutting and splicing. However, larger-scale bilingual datasets covering flexible multi-span instructions, location-reference editing, and paralinguistic editing remain scarce.

Building a large-scale, high-quality dataset is challenging for three reasons.
First, paired speech editing data are scarce and cannot be collected at scale.
Even when one speaker reads both the original and edited sentence, it is hard to keep the two speech samples consistent outside the edited region because their prosody and acoustic environment differ.
Second, constructing paired data by cutting and splicing real speech based on predefined rules avoids collecting new speech \citep{chen2026cosyedit}, but it restricts edits to words that the sample already contains and can introduce abrupt transitions and discontinuous background sound at the edit boundaries.
Finally, constructing high-quality editing pairs through speech synthesis requires accurate edit localization and reliable quality control. Existing large-scale speech datasets often lack word-level timestamps and nonverbal event boundaries, making it difficult to precisely identify the edit regions. Moreover, synthesized speech may contain content errors or speaker inconsistencies, requiring additional verification to ensure the quality of editing pairs.

To address these challenges, we build a scalable pipeline that leverages advanced TTS models and a comprehensive quality control mechanism to construct high-quality speech editing pairs from open-source speech corpora
(Figure~\ref{fig:speech-editing-pipeline}).
Using this pipeline, we construct \dataset{}, an open-source
bilingual speech editing dataset containing approximately
41M audio pairs, with roughly 20.5M pairs in each of Chinese
and English.
\dataset{} covers both content and paralinguistic editing.
Content editing includes single-span editing, which inserts,
deletes, or substitutes words at one location; multi-span
editing, which modifies multiple locations with a single
instruction.
The dataset also includes location-referenced editing that specifies edit positions by ordinal position, time interval,
sentence boundary, or all occurrences of a word.
Paralinguistic editing covers the insertion and deletion of 39 classes of nonverbal vocalizations, including laughter, sighs, and coughs.

To validate the effectiveness of \dataset{}, we fine-tune
AuK \citep{ma2026auk}, a state-of-the-art speech editing
model, using LoRA with only approximately 1.3\% trainable
 parameters relative to the original model.
The fine-tuned model consistently outperforms the original AuK on SpeechEditBench \citep{zhang2026speecheditbench} and Ming-Freeform-Audio-Edit\citep{yan2025muniaudio}, demonstrating the effectiveness of \dataset{} in improving speech editing performance through parameter-efficient fine-tuning.

Beyond evaluating the effectiveness of \dataset{}, we systematically investigate the learning dynamics of speech editing. Our analysis reveals distinct convergence behaviors and interactions among editing operations and examines how data scale and task composition affect their learning. We further show that speech editing and generation provide complementary learning signals, while curriculum learning and data reversal offer effective strategies for improving complex editing and preserving acoustic quality. Finally, we investigate the different learning trajectories of content and paralinguistic editing capabilities. Our contributions are as follows:
\begin{itemize}
    \item We release \dataset{}, to the best of our knowledge,
    the largest publicly available speech editing dataset,
    containing approximately 41M audio pairs in Chinese
    and English. The dataset covers diverse content and
    paralinguistic editing tasks.
    \item We open-source a scalable data pipeline to facilitate the development of speech editing datasets and advance research in this field.
    The pipeline can also serve as a reusable component for broader audio-visual applications, such as dubbing and joint editing.
    \item We demonstrate the effectiveness of \dataset{} by fine-tuning AuK with LoRA, achieving consistent improvements over the original model on SpeechEditBench and Ming-Freeform-Audio-Edit.
    \item We systematically investigate the learning dynamics of speech editing, revealing how different editing capabilities emerge and interact. Our findings provide practical insights into data design and training strategies for developing more effective speech editing models.
\end{itemize}

\section{Related Work}
\label{sec:related}

\subsection{Instruction-based Speech Editing Methods}
Speech editing was first studied in the infilling-based setting, from early splicing and masked-infilling methods \citep{jin2017voco, tan2021editspeech, wang2022campnet, bai2022a3t, jiang2023fluentspeech} to large speech generation models \citep{le2023voicebox, peng2024voicecraft, wang2025ssrspeech, zheng2025voicecraftx, chen2026cosyedit}.
Instruction-based methods replace the edited transcript with an instruction.
InstructSpeech \citep{huang2024instructspeech} follows editing instructions with an LLM, and recent speech generation models follow instructions to edit content \citep{yan2025muniaudio, wang2026dotsttsedit, shen2026fireredtts3, ma2026auk}, emotion and speaking style \citep{yan2025stepaudioeditx, yan2025muniaudio, wang2026dotsttsedit}, and nonverbal vocalizations \citep{yan2025stepaudioeditx, ma2026auk}. None of these instruction-based methods release their training data, so they cannot be retrained or compared on common training data.

Many instruction-based models also support speech editing and generation within a unified framework \citep{yan2025stepaudioeditx, yan2025muniaudio, shen2026fireredtts3, ma2026auk}, building on earlier multi-task speech models \citep{wang2024speechx, yang2024uniaudio}. Recent studies further demonstrate knowledge transfer between these tasks. CosyEdit\citep{chen2026cosyedit} adapts a pretrained TTS model to speech editing, while CosyEdit2\citep{chen2026cosyedit2} shows that editing-oriented reinforcement learning can improve TTS performance. Although task interactions and learning dynamics have been explored in other fields \citep{standley2020tasks, yu2020pcgrad, swayamdipta2020cartography, ren2025learningdynamics}, existing speech editing studies primarily focus on the performance of trained models. How different editing capabilities emerge and interact during training, and how speech generation influences their acquisition, remain largely unexplored.

\subsection{Instruction-based Speech Editing Datasets}
Large-scale paired editing datasets have enabled instruction-based editing in other modalities. For image editing, InstructPix2Pix \citep{brooks2023instructpix2pix} and MagicBrush \citep{zhang2023magicbrush} introduce synthetic and manually annotated editing pairs, while UltraEdit \citep{zhao2024ultraedit} and AnyEdit \citep{yu2025anyedit} scale data construction to millions of samples. For general audio editing, AUDIT \citep{wang2023audit} constructs instruction-based editing pairs from mixtures of sound events.

In speech editing, publicly available datasets primarily serve as evaluation benchmarks. RealEdit \citep{peng2024voicecraft} evaluates infilling-based content editing on real speech, while more recent benchmarks cover content, emotion, style, paralinguistic, and compositional editing \citep{yan2025muniaudio, yan2025stepaudioeditx, wang2026dotsttsedit, zhang2026speecheditbench, ma2026mmae}.

Publicly available training datasets remain limited in scale and task coverage. GigaEdit-S and GigaEdit-L \citep{chen2026cosyedit} provide approximately 0.5M and 2.1M content editing pairs, respectively, constructed through cutting and splicing real speech. Although they support basic content editing and multi-span operations, their construction is restricted to existing speech segments and does not provide natural-language editing instructions. ISSE \citep{chen2025isse} provides over 100K instruction-based speech style editing pairs, while FineEdit \citep{zhou2026fineedit} focuses on prosody, emotion, and timbre control without content editing. Meanwhile, large-scale training datasets used by recent instruction-based speech editing models remain largely in-house \citep{wang2026dotsttsedit, ma2026auk}. Existing nonverbal vocalization datasets provide event annotations for recognition and expressive speech synthesis but do not contain speech editing pairs \citep{gong2022vocalsound, nguyen2023expresso, borisov2025nonverbaltts, liao2025nvspeech}.

Overall, publicly available speech editing training datasets remain limited in scale and diversity, particularly for location-reference editing, flexible multi-span instructions, and paralinguistic editing. To our knowledge, \dataset{} is the first large-scale, high-quality, publicly available bilingual speech editing training dataset covering both content and paralinguistic editing, with natural-language instructions for every editing pair. We also open-source our scalable data construction pipeline, incorporating multi-stage quality filtering and human spot checks to facilitate reliable and reproducible dataset construction.
\section{\dataset{}}
\label{sec:dataset}
Each sample in \dataset{} is represented as an editing
triplet $\langle x_{\mathrm{src}}, c, x_{\mathrm{tgt}}\rangle$,
where $x_{\mathrm{src}}$ denotes the source speech,
$c$ is the editing instruction, and $x_{\mathrm{tgt}}$
is the target edited speech. An overview of our annotation pipeline is
illustrated in Figure~\ref{fig:speech-editing-pipeline}.

\subsection{Editing Taxonomy}
\label{sec:data-tasks}
\dataset{} contains two major categories: content editing and paralinguistic editing.
\paragraph{Content Editing.} Content editing is further divided into three subcategories:
\textbf{(a) Single-span editing} inserts, deletes, or substitutes words at a single location in the speech.
\textbf{(b) Multi-span editing} applies two to six edits within a single instruction, where each edit can independently be an insertion, deletion, or substitution.
\textbf{(c) Location-reference editing} specifies edit locations through occurrence-based references (ordinal positions or all occurrences of a word), sentence-boundary references, or time-based references (time points or intervals).

\paragraph{Paralinguistic Editing.} Paralinguistic editing covers the insertion and deletion of nonverbal vocalizations (e.g., laughter, coughing, and sighing). Table~\ref{tab:subsets} presents example instructions and the number of pairs for each editing subset.

\begin{table*}[t]
  \centering
  \small
  \setlength{\tabcolsep}{5pt}
  \begin{tabular}{@{}lllc@{}}
    \toprule
    Task & English instruction & Chinese instruction & \#Pairs \\
    \midrule

    \rowcolor{gray!12}
    \multicolumn{4}{@{}l}{\emph{Single-span editing}} \\
    \quad Insertion & Add `all' after `you'. & \zh{在‘你们’后面加上‘都’。} & 10M \\
    \quad Deletion & Delete `very' before `good'. & \zh{删除‘好’前面的‘很’。} & 10M \\
    \quad Substitution & Replace `red' with `blue'. & \zh{把‘红色’改成‘蓝色’。} & 10M \\

    \rowcolor{gray!12}
    \multicolumn{4}{@{}l}{\emph{Multi-span editing} (2--6 edits per instruction)} \\
    \multirow{2}{*}{\quad Two edits} & \multicolumn{2}{l}{Add `all' after `you'; delete `very' before `good'.} & \multirow{2}{*}{10M} \\
    & \multicolumn{2}{l}{\zh{在‘你们’后面加上‘都’；删除‘好’前面的‘很’。}} & \\

    \rowcolor{gray!12}
    \multicolumn{4}{@{}l}{\emph{Location-reference editing}} \\
    \quad Ordinal & Delete the second `the'. & \zh{删除第二个‘的’。} & 100K \\
    \quad All occurrences & Replace every `red' with `blue'. & \zh{把所有的‘红色’改成‘蓝色’。} & 100K \\
    \quad Sentence boundary & Delete the sentence-initial `so'. & \zh{删除句首的‘所以’。} & 100K \\
    \quad Time & Delete the speech from 3\,s to 5\,s. & \zh{删除第3秒到第5秒的语音。} & 100K \\

    \rowcolor{gray!12}
    \multicolumn{4}{@{}l}{\emph{Paralinguistic editing}} \\
    \quad Insertion & Add `[surprise-oh]' before `am allowed'. & \zh{在‘像蚊子哼哼’前面加上‘[laugh]’。} & 210K \\
    \quad Deletion & Delete `[surprise-oh]' before `am allowed'. & \zh{删除‘像蚊子哼哼’前面的‘[laugh]’。} & 210K \\

    \midrule
    Total & & & $\sim$41M \\
    \bottomrule
  \end{tabular}
  \caption{Editing tasks in \dataset{}, with example instructions in English and Chinese and the total number of editing pairs across both languages.}
  \label{tab:subsets}
\end{table*}

\subsection{ASR \& Forced Alignment}
\label{sec:data-source}

Both the content and the paralinguistic editing data of \dataset{} are built from open-source speech corpora (Figure~\ref{fig:speech-editing-pipeline}B).
The content editing data come from the Chinese and English subsets of Emilia \citep{he2024emilia} and the paralinguistic editing data are constructed from six existing nonverbal speech corpora \citep{liao2025nvspeech,bai2026synparaspeech,ye2025scalable,SMIIP-NV,mai2026mnv,borisov2025nonverbaltts}.

To obtain high-quality transcriptions and precise word-level timestamps, we re-transcribe all source audio using Qwen3-ASR-1.7B and perform forced alignment using Qwen3-ForcedAligner-0.6B \citep{shi2026qwen3asr} (Figure~\ref{fig:speech-editing-pipeline}C). Additional experiments in Appendix~\ref{app:retranscription} demonstrate the effectiveness of re-transcription in improving speech intelligibility performance.

\subsection{Instruction Generation}
\label{sec:data-instruction}
To construct single-span editing instructions, we first select an edit region using predefined sampling rules. For insertion and substitution, we employ Qwen3.6-35B-A3B \citep{qwen2026qwen36} to generate new content conditioned on the source transcript and selected target (Figure~\ref{fig:speech-editing-pipeline}D). The model is instructed to ensure that the edited transcript remains grammatically fluent and semantically coherent. For deletion, the selected content is removed directly. The final editing instructions are assembled using predefined templates. Detailed LLM prompts are provided in Appendix~\ref{app:llm-prompts}.

For multi-span editing, we first randomly sample the number of edits from two to six. The model then generates the specified number of edits in a single pass, with each operation independently selected from insertion, deletion, and substitution. The edited regions are required to be non-overlapping to avoid interference between edits.

To support more flexible and realistic editing requests, we additionally construct location-reference editing data. Edit locations are specified through occurrence-based references (e.g., the second occurrence of a word), sentence-boundary references (e.g., the beginning of a sentence), or time-based references (e.g., the interval from 3 to 5 seconds). 
\begin{figure*}[ht]
    \centering
    \includegraphics[width=\textwidth]{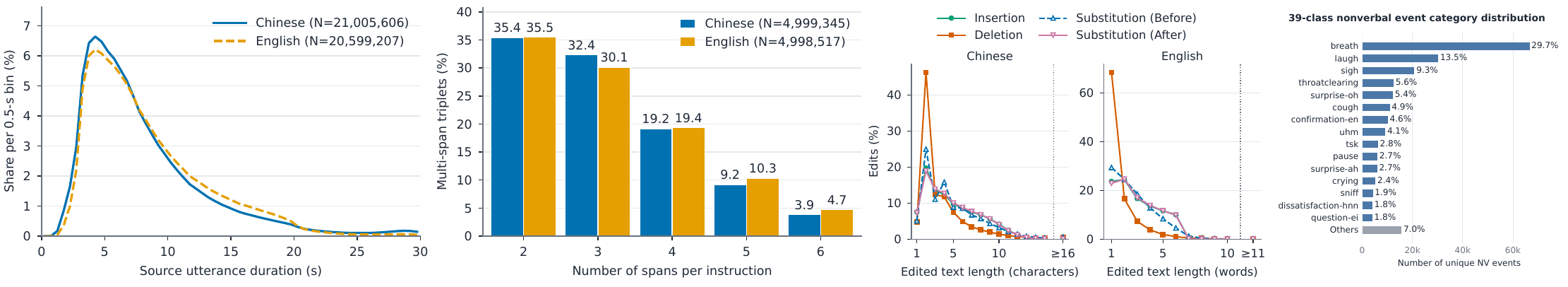}
    \caption{Dataset statistics of \dataset{}. From left to right: source utterance duration distributions, multi-span complexity, edit-span length distributions, and nonverbal event category distributions.}
    \label{fig:data_statistics}
\end{figure*}

\subsection{Edited Paired Speech Generation}
\label{sec:data-synthesis}
To modify only the intended edit regions while preserving the remaining speech, we use the previously obtained word-level timestamps to locate the edited regions. For content editing, these regions are determined by the aligned words or insertion positions. For paralinguistic editing, we infer the locations of nonverbal vocalizations from their annotations and the timestamps of surrounding spoken words. We then employ F5-TTS \citep{chen2025f5tts} to regenerate the masked edited regions conditioned on the edited transcript and the surrounding audio context (Figure~\ref{fig:speech-editing-pipeline}E). To avoid audible discontinuities at the edit boundaries, we extend each masked region by 0.5\,s on both sides, while keeping the audio outside the extended regions unchanged.

To mitigate potential quality degradation caused by synthesized training targets, we adopt a pair reversal strategy inspired by recent image and video editing datasets \citep{winter2024objectdrop,wasserman2024paintbyinpaint,zi2026refvideo}. For single-span, multi-span, and paralinguistic editing, we reverse the editing pairs, using synthesized speech as the source, real speech as the target, and the inverse edit as the instruction. For location-reference editing, we retain the original editing direction because location-dependent instructions cannot always be reliably inverted. We further evaluate the effectiveness of this strategy in preserving speech quality in Section~\ref{sec:analysis-reversal}.
\subsection{Quality Control}
\label{sec:data-qc}
To ensure the quality and reliability of synthesized editing pairs, we apply three automatic filters followed by human spot checks (Figure~\ref{fig:speech-editing-pipeline}F). First, a rule-based filter removes samples with synthesized speech longer than 30\,s. Second, we transcribe the synthesized edited speech using Qwen3-ASR-1.7B and compare their transcription with the expected transcript. For English, we retain only samples with zero word error rate (WER). For Chinese, we convert both transcripts into pinyin sequences and require zero pinyin error rate (PER), avoiding penalties for homophonous character differences in ASR outputs. For nonverbal editing, the expected transcript excludes the annotated nonverbal vocalizations, allowing the same ASR-based filter to verify the preservation of spoken content. Third, we compute the speaker similarity between the original and synthesized speech using WavLM-TDNN \citep{chen2022wavlm} speaker embeddings, retaining only pairs with a similarity above 0.5. Finally, human annotators conduct random spot checks on the remaining pairs.
\subsection{Dataset Statistics}
\label{sec:data-stats}

\dataset{} contains approximately 41M editing pairs, with about 20.5M pairs in each of Chinese and English (examples are in Table~\ref{tab:subsets}). Figure~\ref{fig:data_statistics} presents the distributions of source utterance duration, multi-span complexity, edit-span length, and nonverbal event categories. More details can be found in Appendix~\ref{app:data-stats}.

















\begin{table*}[ht]
\centering
\small
\setlength{\tabcolsep}{4pt}
\renewcommand{\arraystretch}{1.15}
\resizebox{\textwidth}{!}{%
\begin{tabular}{@{}p{0.25\textwidth}lccccc@{}}
\toprule
\textbf{Benchmark} & \textbf{Model} & \textbf{Metrics}
& \multicolumn{4}{c}{\textbf{Results}} \\
\midrule

\multirow{4}{*}{
    \shortstack[l]{
        \textbf{SpeechEditBench}\\
        \textit{Content | Paralinguistic}
    }
}
& & &
\textit{Content-ZH} &
\textit{Content-EN} &
\textit{Avg.} &
\textit{Paralinguistic} \\
\cmidrule(lr){4-7}

& AuK
& \multirow{3}{*}{ACC$\uparrow$}
& 88.32 & 93.78 & 91.05 & 53.25 \\

& AuK + GigaEdit-L
&
& -- & 94.22 & -- & -- \\

& \model{}
&
& \textbf{93.66} & \textbf{96.78}
& \textbf{95.22} & \textbf{54.50} \\

\midrule

\multirow{13}{*}{
    \shortstack[l]{
        \textbf{Ming-Freeform-Audio-Edit}\\
        \textit{Semantic Editing}
    }
}
& & &
\textit{Basic-ZH} &
\textit{Full-ZH} &
\textit{Basic-EN} &
\textit{Full-EN} \\
\cmidrule(lr){4-7}

& AuK
& \multirow{3}{*}{WER$\downarrow$}
& 4.84 & 3.09 & 4.05 & 4.11 \\

& AuK + GigaEdit-L
&
& -- & -- & 5.80 & 5.95 \\

& \model{}
&
& \textbf{4.40} & \textbf{2.93}
& \textbf{3.68} & \textbf{3.77} \\

\cmidrule(lr){2-7}

& AuK
& \multirow{3}{*}{ACC$\uparrow$}
& 92.03 & 91.05 & 85.93 & 85.69 \\

& AuK + GigaEdit-L
&
& -- & -- & 83.94 & 83.50 \\

& \model{}
&
& \textbf{92.62} & \textbf{91.58}
& \textbf{87.51} & \textbf{86.84} \\

\cmidrule(lr){2-7}

& AuK
& \multirow{3}{*}{SIM$\uparrow$}
& 0.87 & 0.88 & 0.88 & 0.88 \\

& AuK + GigaEdit-L
&
& -- & -- & \textbf{0.94} & \textbf{0.94} \\

& \model{}
&
& \textbf{0.89} & \textbf{0.90}
& 0.89 & 0.89 \\

\cmidrule(lr){2-7}

& AuK
& \multirow{3}{*}{
    \shortstack[c]{no-edit\\WER$\downarrow$}
}
& 5.00 & 3.17 & 15.52 & 15.58 \\

& AuK + GigaEdit-L
&
& -- & -- & 16.92 & 17.07 \\

& \model{}
&
& \textbf{4.52} & \textbf{3.05}
& \textbf{15.46} & \textbf{15.57} \\

\bottomrule
\end{tabular}%
}
\caption{Speech editing performance of AuK fine-tuned on different datasets. Avg. denotes the average content editing accuracy across Chinese and English on SpeechEditBench. GigaEdit-L contains only English training data, so its Chinese results and bilingual averages are omitted. Best results are highlighted in bold.}
\label{tab:editing-performance}
\end{table*}

\section{Experiments}
\label{sec:experiments}

\subsection{Speech Editing Model}
\label{sec:model}
To evaluate the effectiveness of \dataset{}, we fine-tune AuK\citep{ma2026auk} with LoRA\citep{hu2021lora} on \dataset{}. We denote the fine-tuned model as \model{}. AuK consists of a frozen Qwen2.5-Omni\citep{xu2025qwen25omni} encoder, an audio VAE, and a flow-matching Transformer. The Qwen2.5-Omni encoder extracts semantic features from the source speech and editing instruction. The VAE encodes the source speech into latents. The flow-matching Transformer combines these features with noisy target latents through dual-stream MMDiT\citep{esser2024sd3} and single-stream DiT\citep{peebles2023dit} blocks. It predicts the flow velocity to generate the edited speech latents. Finally, the VAE decoder reconstructs the edited speech.

\subsection{Evaluation Benchmark and Metrics}
\label{sec:setup}
We evaluate speech editing performance on SpeechEditBench \citep{zhang2026speecheditbench} and Ming-Freeform-Audio-Edit\citep{yan2025muniaudio}, following their respective evaluation protocols. The original AuK serves as the baseline.

For SpeechEditBench, we evaluate the Chinese and English content and paralinguistic editing subsets and report editing accuracy (ACC), corresponding to its joint success metric. For the learning dynamics analysis, we denote content editing ACC as Edit Success (ES) and additionally report Exact Match (EM). EM measures the proportion of samples whose normalized transcriptions exactly match the target transcripts, while ES measures whether the requested edit is successfully performed.

For Ming-Freeform-Audio-Edit, we evaluate semantic editing under the Basic and Full settings in both Chinese and English. We report WER, editing accuracy (ACC), speaker similarity (SIM), and no-edit WER. For ASR-based evaluation, edited speech is transcribed using Whisper-Large-v3 \citep{radford2023whisper} for English and Paraformer \citep{gao2022paraformer} for Chinese. 

For both the main experiments and the learning dynamics analysis, all evaluations are conducted with three different random seeds, and the reported results are averaged across the three runs. Detailed definitions and computation procedures for the evaluation metrics are provided in Appendix~\ref{app:evaluation-metrics}.

\subsection{Implementation Details}
\label{sec:imp}
We fine-tune AuK \citep{ma2026auk} using low-rank adaptation (LoRA) \citep{hu2021lora}, with approximately 20M trainable parameters. The LoRA rank, alpha, and dropout are set to 16, 32, and 0, respectively. LoRA adapters are applied to the attention projections, feed-forward layers, adaptive normalization layers, and output projections. 
Training is conducted for one epoch using AdamW with a learning rate of $5\times10^{-4}$, 500 warmup steps, and a total batch size of 2.24M frames.

For the learning dynamics analysis in Section~\ref{sec:analysis}, we adopt a smaller AuK-based model with approximately 278M trainable backbone parameters. The backbone consists of 4 dual-stream MMDiT blocks and 8 single-stream DiT blocks, with a hidden dimension of 1024 and 16 attention heads. Unless otherwise specified, all experiments use English-only training data, and the backbone is trained from scratch for 200K steps. Detailed model and training configurations are provided in Appendix~\ref{app:probe-config}.

During inference, we follow the AuK recommended settings: EMA weights, 32 Euler steps, a classifier-free guidance (CFG) scale of 2.0, and sway sampling \citep{chen2025f5tts} with a coefficient of $-1$.
\section{Results and Learning Dynamics}
\subsection{Speech Editing Results}
Table~\ref{tab:editing-performance} presents the performance comparison between AuK and \model{}. With only approximately 1.3\% of the original AuK backbone parameters updated, \model{} consistently improves speech editing performance across both benchmarks. On SpeechEditBench, \model{} achieves higher accuracy in both content and paralinguistic editing. On Ming-Freeform-Audio-Edit, it improves editing accuracy and linguistic correctness across all Chinese and English settings, while also achieving better speaker similarity and preservation of unedited content. Compared with fine-tuning on GigaEdit-L under the same setup, \dataset{} yields better content editing performance, further demonstrating its effectiveness as training data.

\subsection{Learning Dynamics of Speech Editing}
\label{sec:analysis}

Beyond evaluating the effectiveness of \dataset{}, we systematically investigate the learning dynamics of speech editing through eight research questions, focusing on task-specific convergence, interactions among editing tasks, data mixture ratios, the relationship between editing and generation, and curriculum learning.

\subsubsection{Effect of Training Data Scale}
\label{sec:analysis-scale}
\paragraph{Q1: Data Scale.} \textit{How does the scale of training data affect the learning dynamics of speech editing?} We investigate three data scales for each fundamental editing operation(insertion, substitution and deletion): 50K, 500K, and 5M samples, approximately 100, 1,000, and 10,000 hours of speech, respectively.

\begin{figure}[t]
    \centering
    \includegraphics[width=\columnwidth]{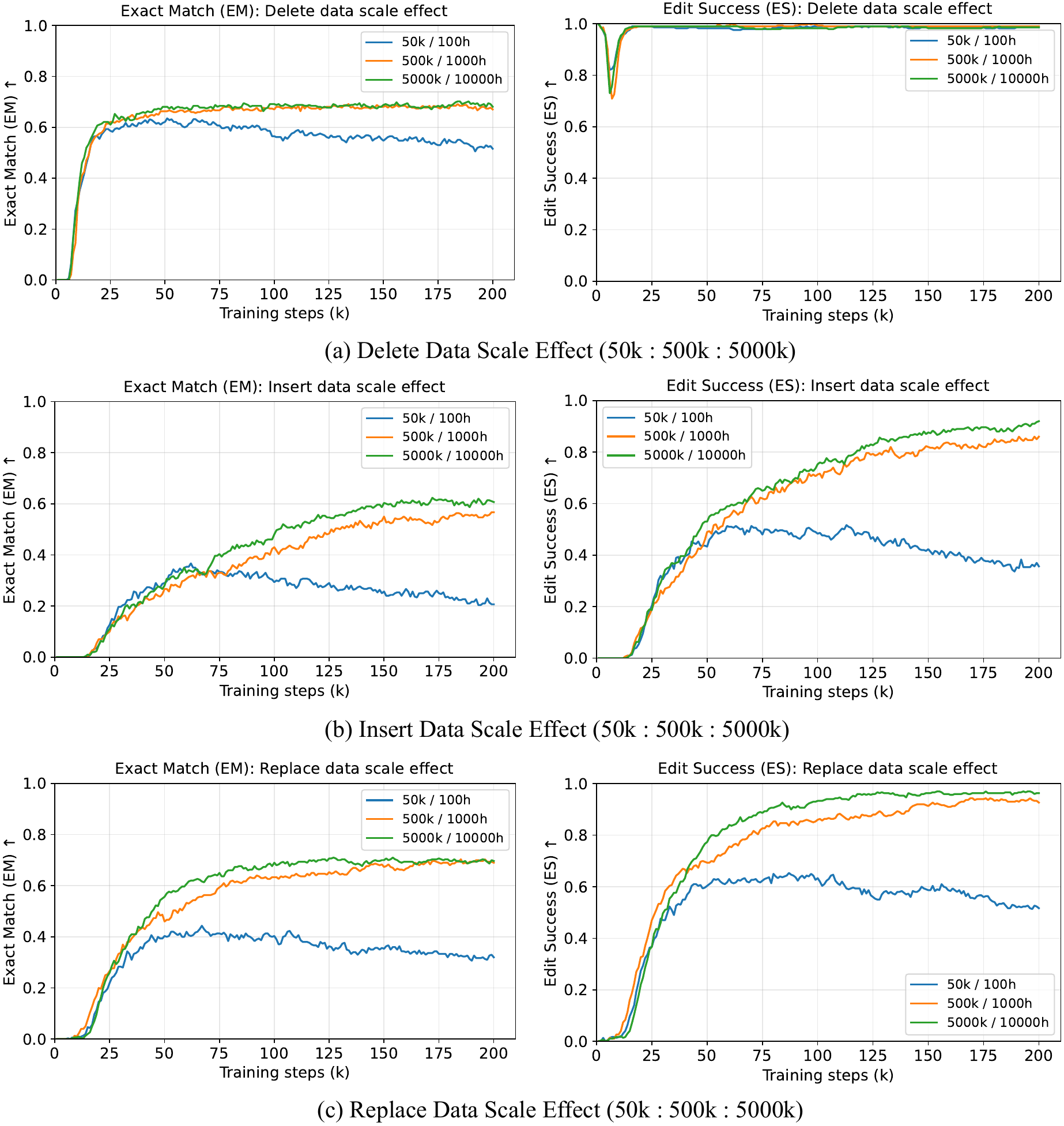}
    \caption{Effect of training data scale on the convergence of insertion, deletion, and substitution.}
    \label{fig:data_scale}
\end{figure}

As shown in Figure~\ref{fig:data_scale}, insertion and substitution benefit substantially from increasing the training data from 50K to 500K samples, while further scaling to 5M yields diminishing returns. In contrast, deletion is less sensitive to data scale and achieves relatively strong performance even with limited training data. Moreover, smaller datasets exhibit performance degradation in later training stages, likely due to overfitting. These findings suggest that editing operations can be categorized into two groups: generation-intensive operations (insertion and substitution), which benefit more from larger datasets, and deletion, which can be learned effectively with much less data.
\subsubsection{Convergence and Task Interactions}
\label{sec:analysis-tasks}

\paragraph{Q2: Task-specific Convergence.} \textit{How do insertion, deletion, and substitution differ in their learning dynamics?} Each editing operation is trained independently to compare its convergence speed and performance.

\begin{figure}[t]
    \centering
    \includegraphics[width=\columnwidth]{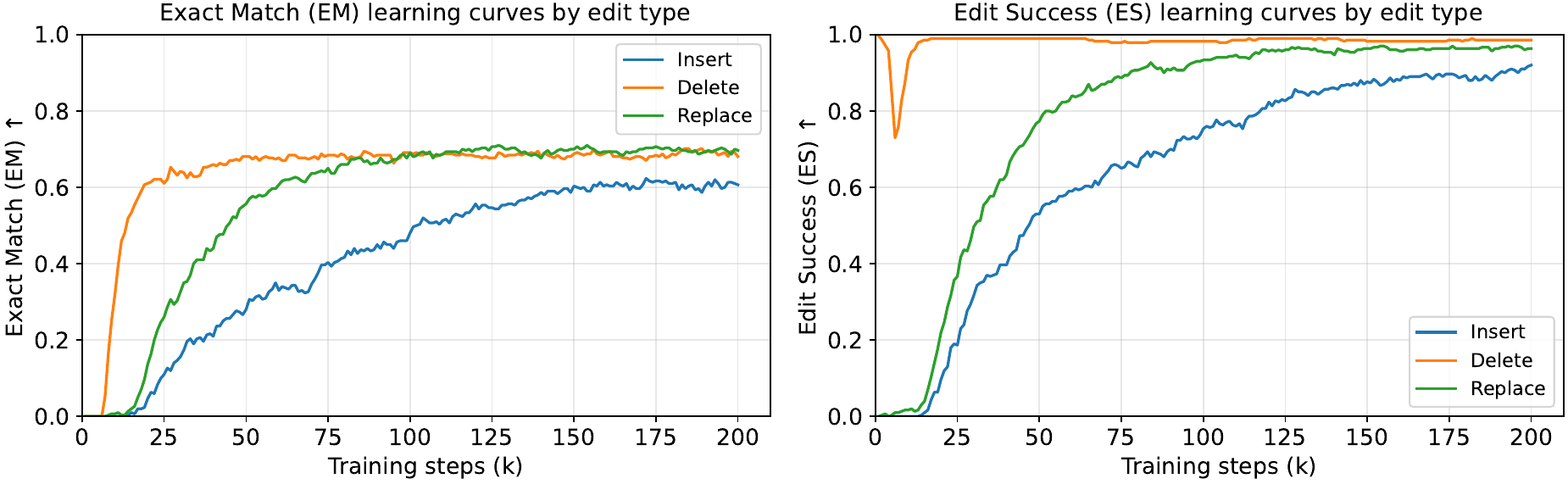}
    \caption{Convergence curves of individual editing tasks under single-operation training.}
    \label{fig:task_convergence}
\end{figure}

As shown in Figure~\ref{fig:task_convergence}, the three operations exhibit distinct convergence behaviors. Deletion converges the fastest, reaching relatively stable performance within approximately 50K steps. In contrast, insertion and substitution require more training steps to learn the generation of new speech content. Insertion converges more slowly than substitution, making it the most challenging operation among the three. These results suggest that editing operations involving new content generation generally require more training than deletion. Similar convergence trends are observed in both Chinese and English, as detailed in Appendix~\ref{app:task-convergence}. Therefore, subsequent learning dynamics analyses focus on English.

\begin{findingbox}
Deletion converges fastest, substitution next, and insertion slowest. Data scaling primarily benefits insertion and substitution, while deletion achieves strong performance with limited data.
\end{findingbox}

\paragraph{Q3: Task Interactions.} \textit{How do insertion, deletion, and substitution interact during joint training?} The experiments compare single-operation training, pairwise training, and joint training of all three editing operations.

\begin{figure}[ht]
    \centering
    \includegraphics[width=\columnwidth]{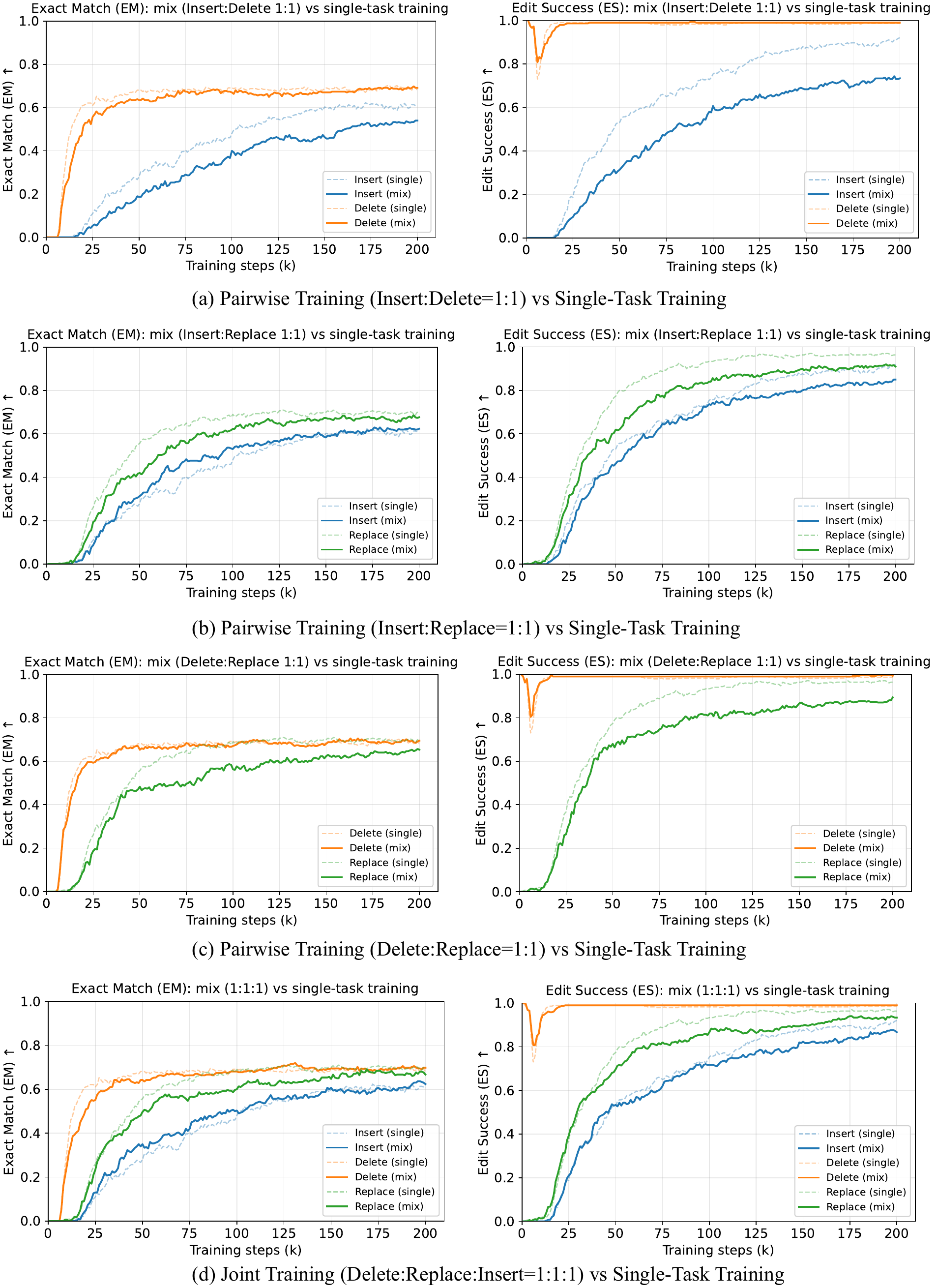}
    \caption{Task interactions under pairwise and joint training. (a)--(c) compare pairwise training with single-operation baselines, while (d) shows joint training of all three operations. Solid and dashed lines represent mixed and single-operation training, respectively.}
    \label{fig:task_interactions}
\end{figure}

\paragraph{Pairwise Training.} As shown in Figure~\ref{fig:task_interactions}(a)--(c), the three operations exhibit asymmetric interactions. Combining insertion with deletion substantially slows insertion learning, while deletion remains largely unaffected. A similar slowdown in substitution is observed when paired with deletion. However, combining insertion with substitution accelerates the early learning of insertion, although substitution converges more slowly. These results suggest that deletion is relatively robust to task mixing, whereas insertion and substitution are more sensitive to the accompanying task.

\paragraph{Joint Training.} Interestingly, the degradation of insertion observed under insertion--deletion training largely disappears when all three operations are jointly trained. Insertion achieves comparable or better performance than single-operation training, while deletion remains stable and substitution shows only a modest decline. This suggests that task interactions depend on the overall task composition, and unfavorable pairwise combinations do not necessarily lead to worse performance when additional tasks are introduced.

\begin{findingbox}
Task interactions depend on task composition. Adding substitution largely alleviates the insertion slowdown observed in insertion--deletion joint training.
\end{findingbox}

\paragraph{Q4: Data Mixture Ratios.} \textit{How does the data balance between generation-intensive operations and deletion affect editing performance?} Building on the preceding findings, we investigate the effects of varying the training data proportions between generation-intensive operations (insertion and substitution) and deletion. Specifically, we compare generation-intensive-to-deletion ratios of 1:1 (1:1:2), 2:1 (1:1:1), and 10:1 (5:5:1), where the ratios in parentheses correspond to insertion, substitution, and deletion, respectively.

\begin{figure}[t]
    \centering
    \includegraphics[width=\columnwidth]{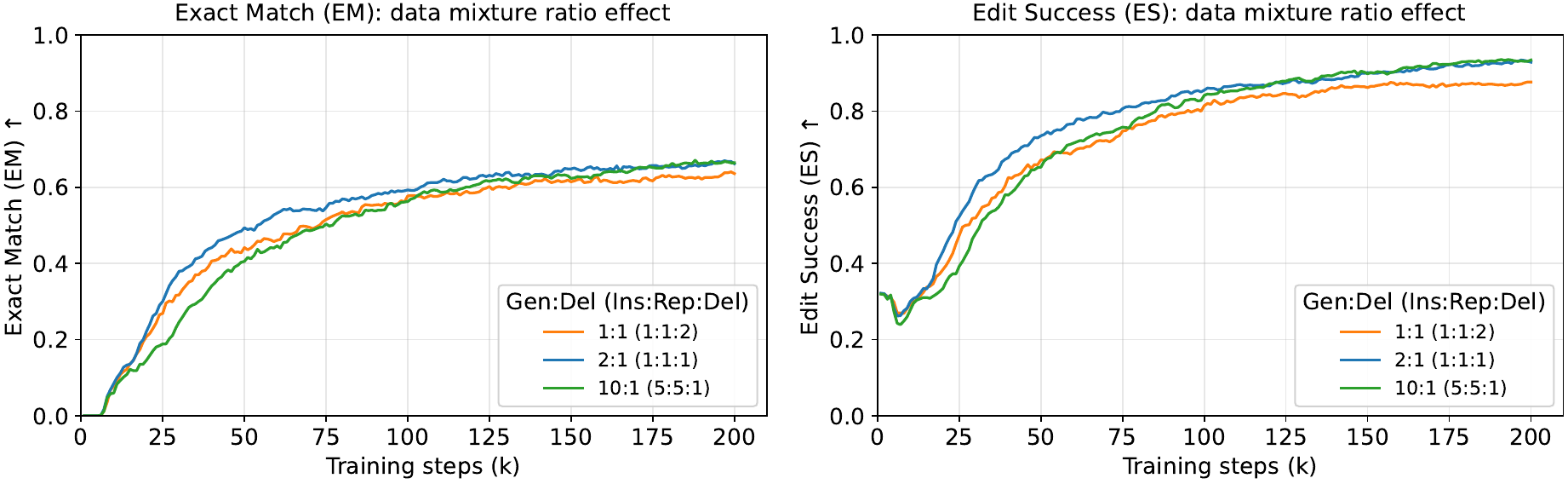}
    \caption{Effect of data mixture ratios on speech editing convergence. The left and right panels show Exact Match (EM) and Edit Success (ES), respectively. The ratios represent generation-intensive operations to deletion, with the corresponding insertion:substitution:deletion ratios shown in parentheses.}
    \label{fig:data_mix_ratio}
\end{figure}

As shown in Figure~\ref{fig:data_mix_ratio}, the 2:1 mixture, which allocates equal training data to the three editing operations, achieves faster convergence and competitive final performance. Increasing the proportion of deletion reduces the performance of generation-intensive operations. Conversely, allocating substantially more data to generation-intensive operations (10:1) yields modest improvements in their final performance, at the cost of deletion performance and slower early convergence. These findings highlight the trade-off between the two operation groups, suggesting that balanced training across individual editing operations provides a favorable combination of convergence efficiency and overall performance. Detailed learning curves are provided in Appendix~\ref{app:mixture-ratios}.

\begin{findingbox}
Balanced training across insertion, deletion, and substitution provides a favorable trade-off between convergence and final performance among the tested mixtures.
\end{findingbox}

\subsubsection{Editing and Generation}
\label{sec:analysis-generation}

\paragraph{Q5: Editing vs. Generation.} \textit{How do speech editing and generation influence each other during training?} We investigate their relationship through two strategies: (i) joint training on editing and generation data and (ii) initializing editing models from pretrained TTS checkpoints.

\paragraph{Edit and Generation Joint Training.} We investigate the effects of joint training on both editing and generation. For editing, we compare editing-only training with editing-to-generation ratios of 1:1 and 1:10. For generation, we compare three configurations: generation-only training, joint training with a 1:1 ratio, and training with the same editing pairs treated as additional generation data. The last configuration controls for the effect of additional training data. 
The generation task follows the zero-shot TTS training formulation of AuK \citep{ma2026auk}. We evaluate generation performance on the Seed-TTS test-en\citep{anastassiou2024seedtts} set following its evaluation protocol. Each checkpoint is evaluated with three random seeds, and the results are averaged.

\begin{figure}[t]
    \centering
    \includegraphics[width=\columnwidth]{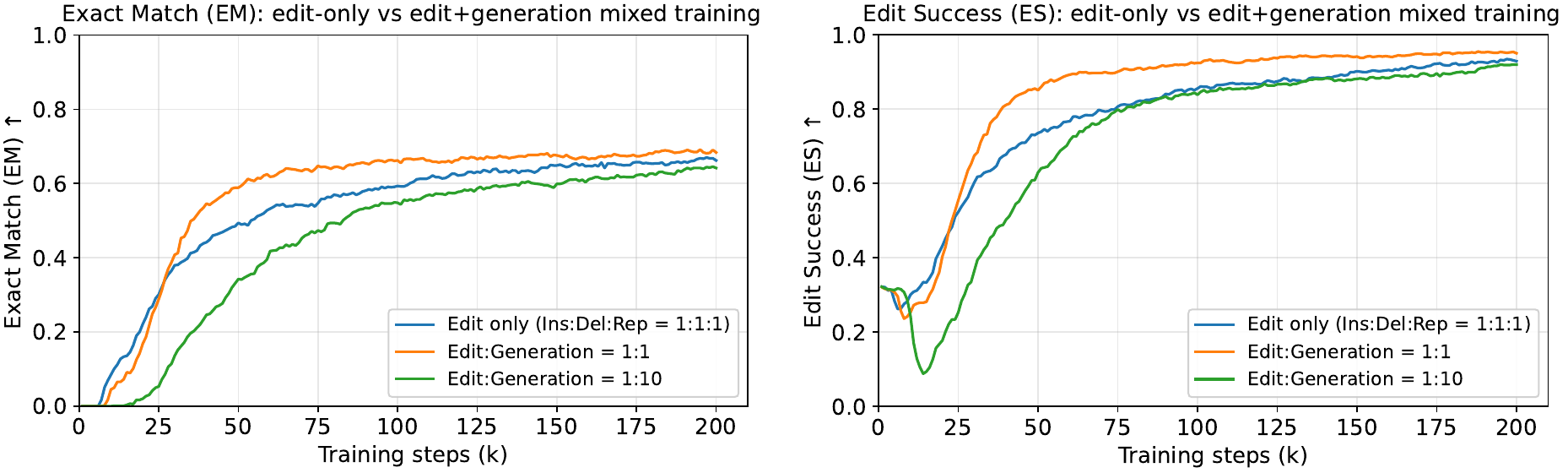}
    \caption{Effect of joint editing and generation training on editing convergence. The left and right panels show Exact Match (EM) and Edit Success (ES), respectively.}
    \label{fig:edit_tts_ratio}
\end{figure}

\textbf{Effect on Editing.} As shown in Figure~\ref{fig:edit_tts_ratio}, joint training substantially accelerates editing convergence, particularly for insertion. This suggests that the speech generation task provides transferable capabilities for editing operations that require synthesizing new content. However, excessive generation data reduces editing performance, indicating that balancing generation and editing supervision is important for effective joint training.

\begin{figure}[t]
    \centering
    \includegraphics[width=\columnwidth]{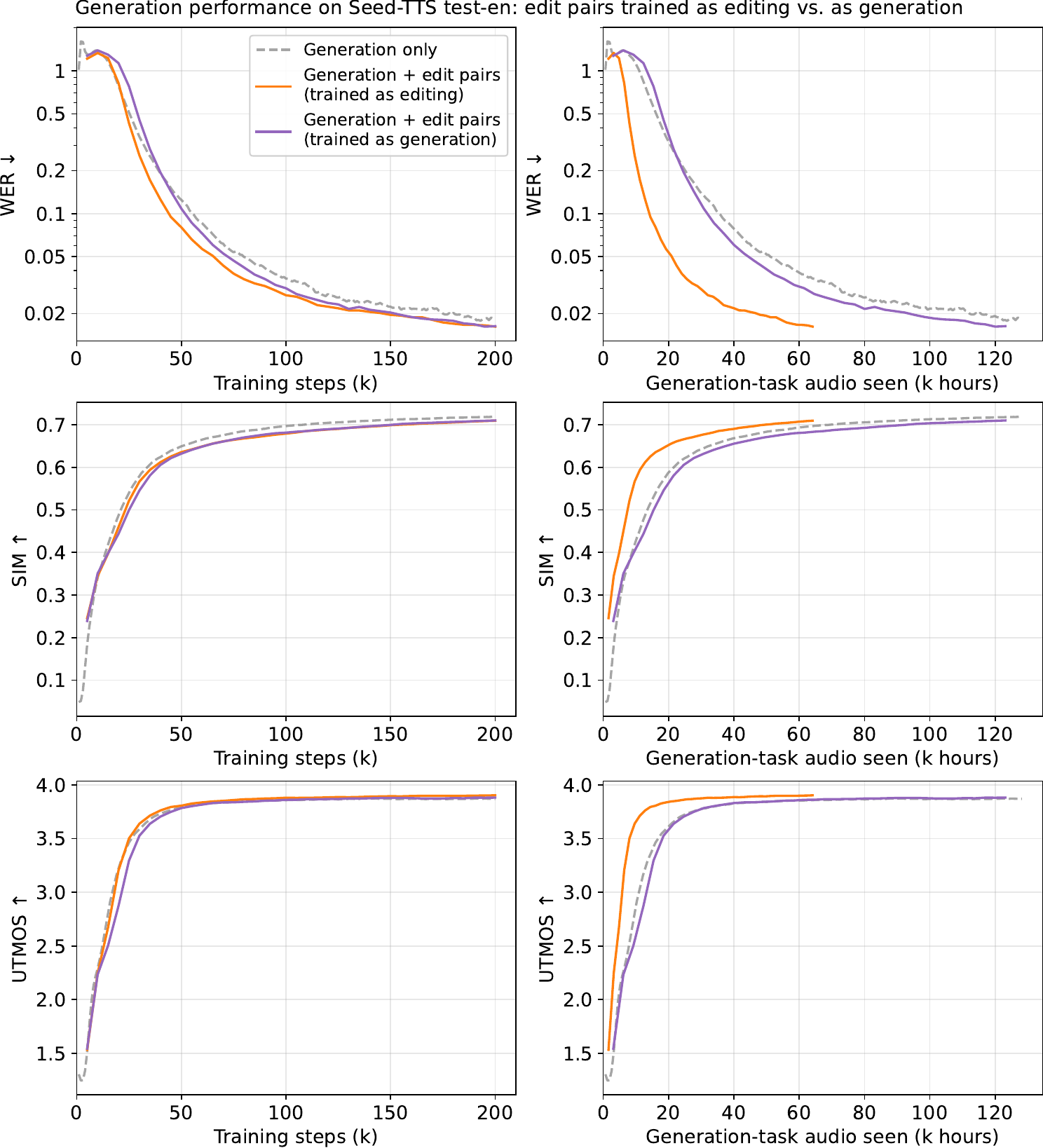}
    \caption{Effect of editing supervision on generation performance on Seed-TTS test-en, measured against training steps (left) and cumulative generation-task audio exposure (right).}
    \label{fig:edit_tts_generation}
\end{figure}

\textbf{Effect on Generation.} As shown in Figure~\ref{fig:edit_tts_generation}, joint training of editing and generation improves generation performance and learning efficiency on the Seed-TTS test set compared with training on generation data alone. To investigate this effect, we compare generation performance under two conditions: \textit{(i) equal training steps} and \textit{(ii) equal generation data budget}, measured by the cumulative hours of generation-task audio seen during training. We consider three training strategies: training on generation data alone, joint training of editing and generation, and treating the additional editing pairs as generation data. The third strategy controls for the effect of data diversity introduced by the editing pairs. Under equal training steps, joint training achieves faster convergence while maintaining comparable speaker similarity and speech naturalness. Under an equal generation data budget, joint training achieves faster improvements across all three metrics. Compared with treating the same editing pairs as generation data, joint training achieves faster convergence, indicating that its benefits extend beyond additional data diversity. More broadly, speech editing can be viewed as an \emph{understanding-grounded generation} task, as it involves recognizing and grounding content
in source speech. These results suggest that understanding-grounded generation tasks can facilitate speech generation.

\begin{figure}[t]
    \centering
    \includegraphics[width=\columnwidth]{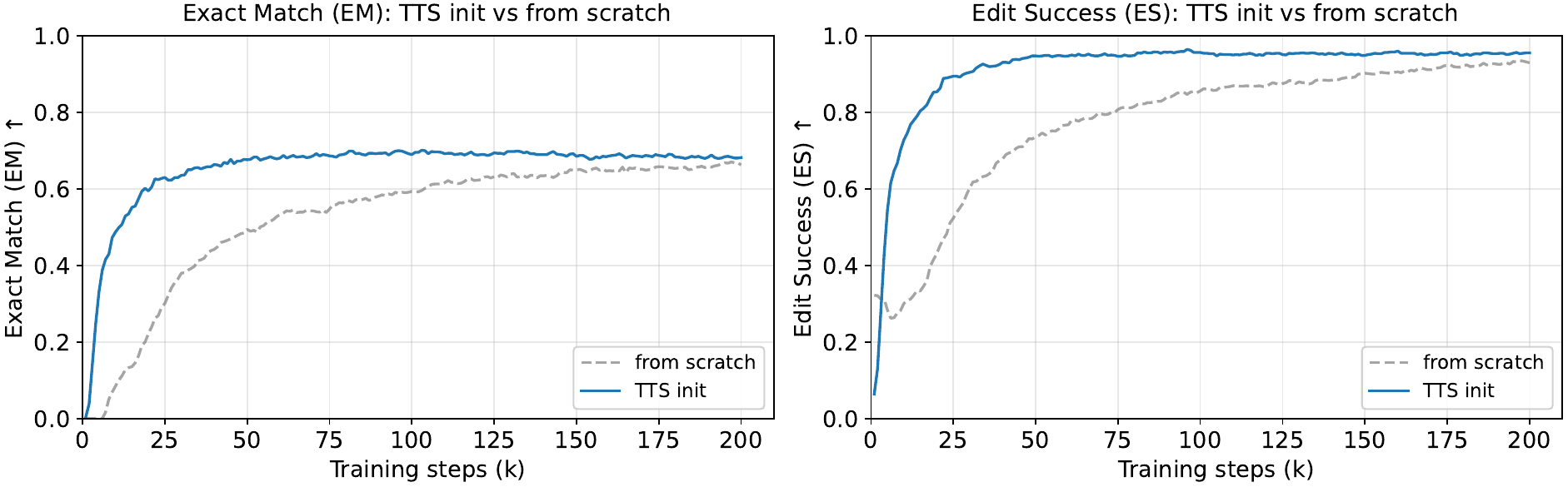}
    \caption{Effect of TTS initialization on speech editing convergence. Overall Exact Match (EM) and Edit Success (ES) are shown on the left and right, respectively.}
    \label{fig:tts_initialization}
\end{figure}

\begin{findingbox}
Editing can be viewed as an understanding-grounded generation task that facilitates speech generation. Joint training improves generation learning efficiency while maintaining comparable generation quality.
\end{findingbox}
\paragraph{Generation Initialization.} As shown in Figure~\ref{fig:tts_initialization}, initializing from a pretrained TTS checkpoint substantially accelerates editing convergence, particularly for insertion and substitution. This improvement may arise from the shared speech synthesis requirements of both tasks. Speech editing requires preserving most of the original content while generating new speech in modified regions. TTS pretraining provides a strong generation prior, allowing the model to focus on learning editing-specific capabilities. Although training from scratch eventually achieves comparable performance, these results demonstrate the effectiveness of transferring speech generation capabilities to editing. Detailed task-level results are provided in Appendix~\ref{app:tts_initialization}.

\begin{findingbox}
Generation initialization can accelerate editing convergence, particularly for insertion and substitution.
\end{findingbox}
\subsubsection{Curriculum Learning}
\label{sec:analysis-curriculum}

\paragraph{Q6: Curriculum Learning.} \textit{Does learning single-span editing before multi-span editing improve editing performance?} We compare three training strategies:(i) \textbf{Single-to-Multi}, which trains on multi-span data starting from a checkpoint pretrained on single-span editing;(ii) \textbf{Multi-only}, which trains on multi-span data from scratch; and (iii) \textbf{Joint Training}, which jointly trains on single-span and multi-span data with a 1:1 ratio from scratch. To evaluate multi-span editing performance, we constructed an evaluation set of 1,000 samples, with 200 samples for each edit count from 2 to 6. 
In addition to the metrics used in SpeechEditBench, we also introduce sub-edit success (subES), which measures the average proportion of successfully executed sub-edits per sample. To reduce transcription errors, we use Qwen3-ASR-1.7B for evaluation.

\begin{figure}[t]
    \centering
    \includegraphics[width=\columnwidth]{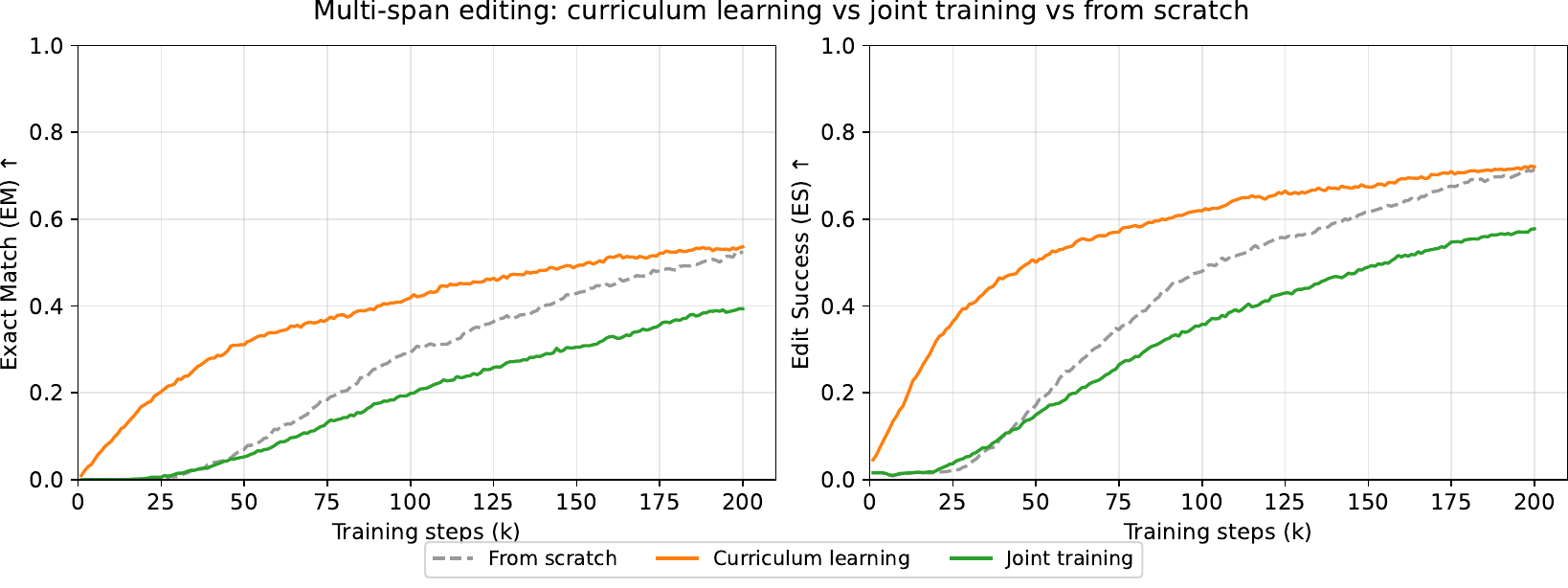}
    \caption{Effect of curriculum learning on multi-span editing convergence. Exact Match (EM) and Edit Success (ES) are shown on the left and right, respectively.}
    \label{fig:curriculum_learning}
\end{figure}

As shown in Figure~\ref{fig:curriculum_learning}, Single-to-Multi substantially accelerates convergence and achieves the best overall performance, while joint training converges more slowly than multi-span-only training. These results suggest that learning fundamental single-span editing operations provides a strong initialization for multi-span editing, allowing the model to acquire combinations of previously learned operations more efficiently. Furthermore, we also observe that the success rate decreases as the number of editing operations increases, highlighting the difficulty of multi-span editing. Detailed results across different numbers of edits are provided in Appendix~\ref{app:curriculum-learning}.

\begin{findingbox}
A simple-to-complex curriculum facilitates multi-span editing. Learning basic single-span operations accelerates convergence and improves overall editing performance.
\end{findingbox}
\subsubsection{Effect of Data Reversal}
\label{sec:analysis-reversal}

\paragraph{Q7: Data Reversal.} \textit{How does data reversal affect speech quality in speech editing?} We compare three training strategies: (i) \textbf{Forward}, which predicts synthetic speech from real speech; (ii) \textbf{Reverse}, which predicts real speech from synthetic speech; and (iii) \textbf{Mixed}, which combines both directions with a 1:1 ratio. Following the findings in Section~\ref{sec:analysis-generation}, all models are initialized from the same pretrained TTS checkpoint and trained for 50K steps on a balanced mixture of insertion, deletion, and substitution. We evaluate speech quality using UTMOS and speaker similarity (SIM) to measure naturalness and speaker consistency.

\begin{figure}[t]
    \centering
    \includegraphics[width=\columnwidth]{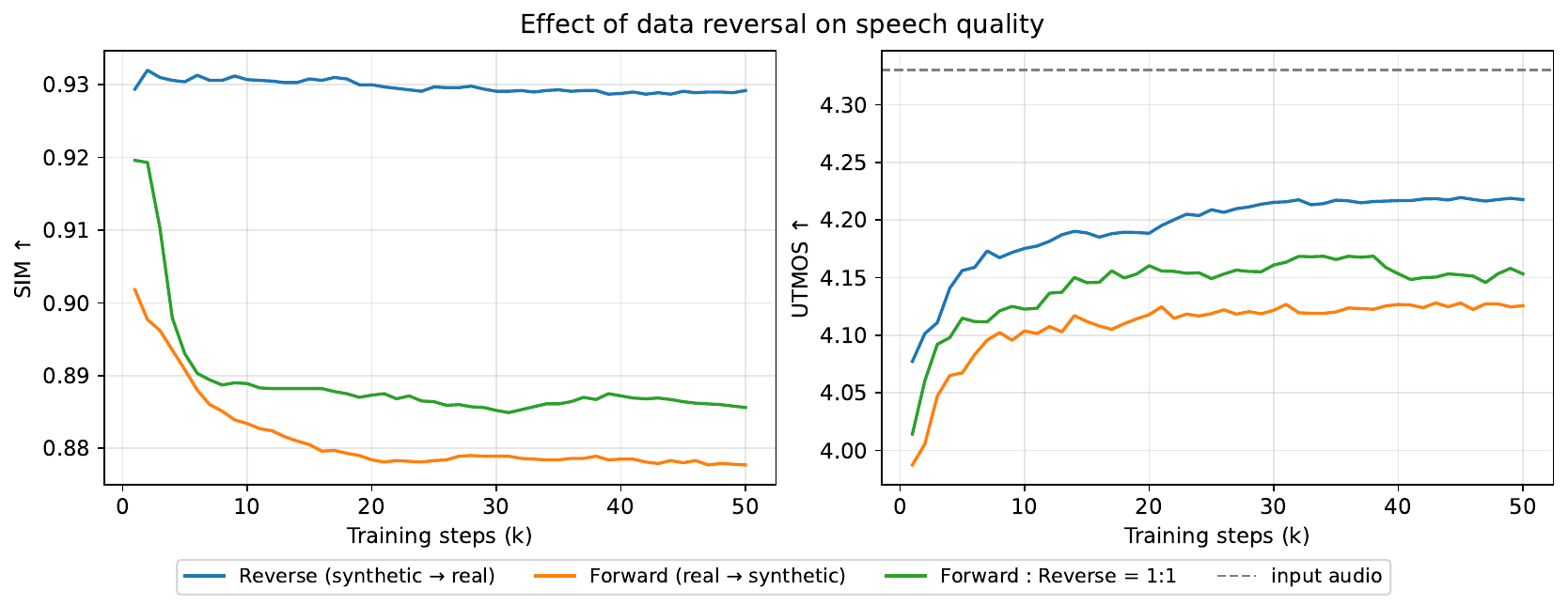}
    \caption{Effect of data reversal on speech quality. Speaker similarity (SIM) and UTMOS are shown on the left and right, respectively. }
    \label{fig:data_reversal}
\end{figure}

As shown in Figure~\ref{fig:data_reversal}, Reverse consistently achieves higher SIM and UTMOS than Forward and Mixed training. This suggests that synthetic targets may introduce acoustic artifacts into the learned generation process, while using real recordings as targets provides cleaner supervision and helps preserve speech quality. These findings support data reversal as an effective strategy for mitigating quality degradation in synthetic speech editing datasets.

\begin{findingbox}
Reversing editing pairs mitigates acoustic quality degradation caused by synthetic training targets, helping preserve speaker similarity and speech naturalness.
\end{findingbox}
\subsubsection{Content vs. Paralinguistic Editing.}
\label{sec:analysis-emergence}
\paragraph{Content and Paralinguistic Editing Learning Progression.}
\textit{How do content and paralinguistic editing capabilities differ in their learning dynamics?}
We train Chinese content and paralinguistic editing models separately for insertion and deletion, and compare their learning dynamics at equal training steps. For paralinguistic editing, we select four common vocalizations (breath, laugh, sigh, and cough), which account for nearly 60\% of the nonverbal events in our released dataset. We construct a paired evaluation set of 100 utterances to minimize the influence of edit localization. For each utterance, content and paralinguistic edits share the same location anchor. A paired editing example is shown below.

\begin{center}
\includegraphics[width=\columnwidth]{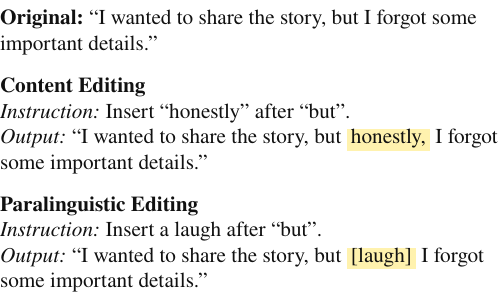}
\end{center}

\begin{figure}[t]
    \centering
    \includegraphics[width=\columnwidth]{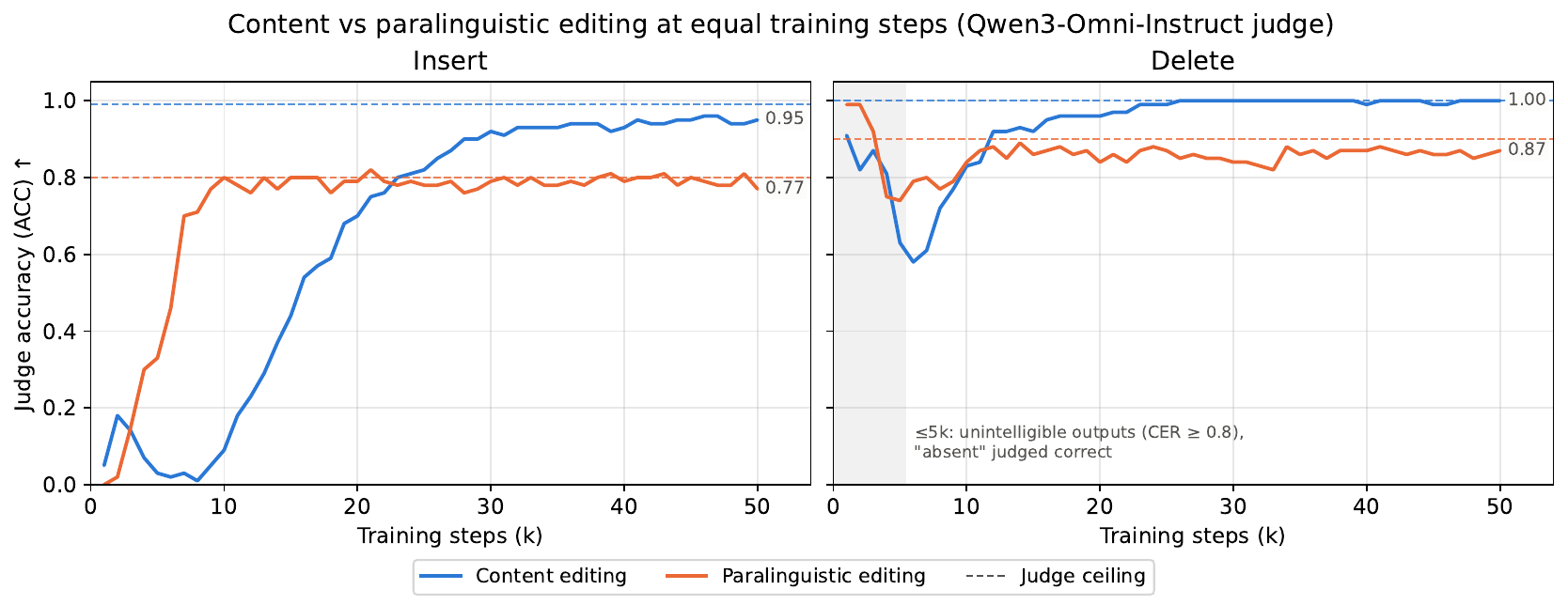}
    \caption{Learning dynamics of content and paralinguistic editing for insertion (left) and deletion (right).}
    \label{fig:nv_content}
\end{figure}

Existing ASR systems struggle to reliably recognize nonverbal vocalizations, making paralinguistic editing difficult to evaluate. To address this limitation, we formulate editing evaluation as a binary classification task following the LLM-as-a-Judge paradigm. Specifically, we use Qwen3-Omni-Instruct\citep{Qwen3-Omni} to determine whether the target word or paralinguistic event is present at the specified location. We also evaluate the judge on the reference target audio, with the resulting accuracies shown as dashed lines in Figure~\ref{fig:nv_content}.

As shown in Figure~\ref{fig:nv_content}, linguistic and paralinguistic editing capabilities exhibit different learning trajectories. For insertion, paralinguistic editing develops faster, showing improvements shortly after the model learns to generate intelligible speech. In contrast, content insertion requires substantially more training, as the model initially tends to reproduce the original speech without inserting new words. For deletion, both capabilities follow similar learning trajectories once the generated speech becomes intelligible. These findings suggest that models may learn to manipulate acoustic events faster than linguistic content, particularly for insertion.

\begin{findingbox}
Paralinguistic insertion develops faster than content insertion, while both deletion capabilities exhibit similar learning trajectories once the generated speech becomes intelligible.
\end{findingbox}
\section{Discussion}
\label{sec:discussion}
\paragraph{Is Multi-turn Editing Necessary for Content Editing?}
Multi-turn editing enables users to iteratively refine speech through successive instructions. However, for content editing involving insertion, deletion, and substitution, multiple editing instructions can often be consolidated into a single instruction covering one or more spans. This raises the question of whether dedicated multi-turn training data are necessary for content editing.

The main challenge arises when instructions depend on editing history, such as reverting a previous edit or reusing an earlier operation. Such cases require tracking previous instructions and intermediate versions. A potential solution is to employ an LLM-based instruction planner that leverages the editing history to resolve these dependencies and converts multi-turn requests into explicit single-turn editing instructions. Therefore, \dataset{} focuses on single-span and multi-span editing, while the benefits of explicit multi-turn training remain an open question.
\section{Conclusion}
In this paper, we introduce \dataset{}, a large-scale bilingual dataset for instruction-based speech editing, along with a scalable data construction pipeline. The dataset covers diverse content and paralinguistic editing tasks. We further train \model{} on \dataset{} to demonstrate its effectiveness. Beyond dataset construction, we systematically investigate the learning dynamics of speech editing. Our experiments reveal that different editing operations exhibit distinct convergence behaviors, with deletion converging faster than insertion and substitution. We also observe asymmetric interactions among editing tasks during joint training and find that incorporating text-to-speech generation can accelerate the learning of editing capabilities. We hope that \dataset{} and these findings will facilitate future research on speech editing.

\section*{Limitations}
While \dataset{} covers content editing and paralinguistic editing in Chinese and English, it does not yet include more diverse paralinguistic editing tasks, such as emotion and speaking style modification, or history-dependent multi-turn editing. Extending the dataset to these scenarios remains a direction for future work.

\bibliography{refs}

\appendix

\section{LLM Prompts for Content Editing}
\label{app:llm-prompts}

This section presents the LLM prompts used for insertion and substitution in content editing. The target word or phrase is selected in advance using predefined rules. For insertion, separate prompts are used to generate content before or after the target. The LLM generates only the inserted or replacement content, while all deletion operations are performed without LLM intervention. The final editing instructions are assembled using predefined templates.

The placeholders \texttt{\{raw\_text\}} and \texttt{\{target\}} denote the original transcript and the selected target word or phrase, respectively. The placeholder \texttt{\{length\_hint\}} provides soft guidance on the desired length of the generated content. For substitution, the target length is sampled by multiplying the original span length by a random factor between 0.5 and 3.0. For insertion, the target length is randomly sampled between 1 and 20. Lengths are measured in Chinese characters or English words, respectively. For location-reference insertion, \texttt{\{direction\}} specifies whether the content is inserted before or after the target, while \texttt{\{direction\_zh\}} provides the corresponding Chinese direction.

\newtcblisting{llmprompt}[1]{
    enhanced,
    breakable,
    listing only,
    colback=black!2,
    colframe=black!60,
    boxrule=0.5pt,
    arc=1mm,
    title={#1},
    fonttitle=\bfseries,
    listing options={
        basicstyle=\ttfamily\scriptsize,
        breaklines=true,
        breakatwhitespace=true,
        columns=fullflexible,
        keepspaces=true,
        showstringspaces=false
    }
}

\subsection{Single-span Editing Prompts}

For single-span editing, we use separate prompts for substitution, insertion before the target, and insertion after the target.

\paragraph{Substitution.}
The substitution prompt generates a replacement phrase for a selected target span.

\begin{llmprompt}{Substitution Prompt}
# Task
You are given raw_text and a target word/phrase
that appears in it. Output exactly ONE new
word/phrase that will replace target.

# Output Language
- The replacement MUST be in the same language
  as target.

# Replacement Rules
- The replacement must be non-empty, pronounceable,
  and different from target.
- It must produce a clearly audible pronunciation
  change when spoken.
- Avoid homophones, near-homophones, spelling-only,
  case-only, and written-form variants.
- Prefer changing an entity, number, object,
  action, or attribute.
- The replacement MUST NOT already appear
  anywhere in raw_text.
- Replacing target must yield grammatically
  fluent and semantically coherent text.
- Do not include leading or trailing punctuation,
  symbols, or whitespace.

{length_hint}

# Input
raw_text: {raw_text}
target: {target}

# Output
Output ONLY the replacement word/phrase as a
single line of plain text.
Do not output explanations, reasoning, labels,
quotes, JSON, or the complete sentence.
\end{llmprompt}

\paragraph{Insertion.}
Insertion uses two separate prompts for inserting content before or after the selected target. Both require the inserted phrase to be fluent, pronounceable, and unambiguously removable.

\begin{llmprompt}{Insert-Before Prompt}
# Task
You are given raw_text and a target word/phrase
that appears in it. Output exactly ONE short
new word/phrase that will be inserted
immediately BEFORE target.

# Direction
The inserted phrase will appear BEFORE target.
The resulting text must be grammatically fluent.

# Output Language
- The inserted phrase MUST be in the same
  language as target.

# Insertion Rules
- Generate non-empty, pronounceable content.
- Keep the insertion a meaningful short fragment.
- Do not generate a complete clause or sentence.
- Avoid unfinished fragments or dangling endings.
- The inserted phrase must read naturally
  when placed before target.
- The phrase MUST NOT already occur immediately
  BEFORE target anywhere in raw_text.
- The phrase MUST NOT contain target
  as a substring.
- Removing the inserted phrase must restore
  the original raw_text exactly.
- Do not include leading or trailing punctuation,
  symbols, or whitespace.

{length_hint}

# Input
raw_text: {raw_text}
target: {target}

# Output
Output ONLY the inserted word/phrase as a
single line of plain text.
Do not output explanations, reasoning, labels,
quotes, JSON, or the complete sentence.
\end{llmprompt}

\begin{llmprompt}{Insert-After Prompt}
# Task
You are given raw_text and a target word/phrase
that appears in it. Output exactly ONE short
new word/phrase that will be inserted
immediately AFTER target.

# Direction
The inserted phrase will appear AFTER target.
The resulting text must be grammatically fluent.

# Output Language
- The inserted phrase MUST be in the same
  language as target.

# Insertion Rules
- Generate non-empty, pronounceable content.
- Keep the insertion a meaningful short fragment.
- Do not generate a complete clause or sentence.
- Avoid unfinished fragments or dangling endings.
- The inserted phrase must read naturally
  when placed after target.
- The phrase MUST NOT already occur immediately
  AFTER target anywhere in raw_text.
- The phrase MUST NOT contain target
  as a substring.
- Removing the inserted phrase must restore
  the original raw_text exactly.
- Do not include leading or trailing punctuation,
  symbols, or whitespace.

{length_hint}

# Input
raw_text: {raw_text}
target: {target}

# Output
Output ONLY the inserted word/phrase as a
single line of plain text.
Do not output explanations, reasoning, labels,
quotes, JSON, or the complete sentence.
\end{llmprompt}

\subsection{Location-reference Editing Prompts}

For occurrence-based location-reference editing, the target word or phrase is selected before LLM inference. The LLM generates only the replacement or insertion content. The edits are subsequently applied to the specified occurrences, and the final instructions are assembled using predefined templates.

\paragraph{Global Substitution.}
For global substitution, the LLM generates a single replacement phrase that is applied to every occurrence of the target expression.

\begin{llmprompt}{Global Substitution Prompt}
# Task
You are given raw_text and a target word/phrase
that appears in it. Output exactly ONE new
word/phrase that will replace EVERY occurrence
of target.

# Output Language
- The replacement MUST be in the same language
  as target.

# Replacement Rules
- The replacement must be non-empty,
  pronounceable, and different from target.
- It must produce a clearly audible
  pronunciation change when spoken.
- Avoid homophones, near-homophones,
  spelling-only, case-only, and written-form
  variants.
- Prefer changing an entity, number, object,
  action, or attribute.
- Keep the replacement length roughly
  comparable to target.
- The replacement MUST NOT already appear
  anywhere in raw_text.
- Replacing EVERY occurrence of target must
  yield fluent and natural text.
- Do not include leading or trailing punctuation,
  symbols, or whitespace.

# Input
raw_text: {raw_text}
target: {target}

# Output
Output ONLY the replacement word/phrase as a
single line of plain text.
Do not output explanations, reasoning, labels,
quotes, JSON, or the complete sentence.
\end{llmprompt}

\paragraph{Location-reference Insertion.}
For occurrence-based insertion, a unified prompt generates a phrase to be inserted before or after the selected target. The insertion direction is specified by the input placeholders. The generated phrase must remain appropriate at the designated occurrences.

\begin{llmprompt}{Location-reference Insertion Prompt}
# Task
You are given raw_text, a target word/phrase,
and an insertion direction.
Output exactly ONE short new word/phrase
to insert {direction} target ({direction_zh}).

# Direction
Insert the new phrase immediately {direction}
target. The resulting text must be fluent.

# Output Language
- The inserted phrase MUST be in the same
  language as target.

# Insertion Rules
- Generate non-empty, pronounceable content.
- Prefer 2-6 Chinese characters or
  1-4 English words.
- Do not generate a whole clause or sentence.
- Avoid unfinished fragments and dangling
  endings.
- The inserted phrase must read naturally
  when placed {direction} target, including
  at repeated occurrences.
- The phrase MUST NOT already occur
  immediately {direction} target in raw_text.
- The phrase MUST NOT contain target
  as a substring.
- Removing the inserted phrase must restore
  the original raw_text exactly.
- Do not include leading or trailing punctuation,
  symbols, or whitespace.

# Input
raw_text: {raw_text}
target: {target}
direction: {direction}

# Output
Output ONLY the inserted word/phrase as a
single line of plain text.
Do not output explanations, reasoning, labels,
quotes, JSON, or the complete sentence.
\end{llmprompt}

\section{Detailed Evaluation Metrics}
\label{app:evaluation-metrics}

This section provides additional details on the evaluation metrics used in SpeechEditBench \citep{zhang2026speecheditbench} and Ming-Freeform-Audio-Edit \citep{yan2025muniaudio}.

\subsection{SpeechEditBench}

\paragraph{Content Editing.}
Exact Match (EM) requires the normalized ASR transcription to exactly match the expected target transcript. Edit Success (ES) evaluates whether the requested operation is successfully performed, without requiring an exact match of the entire transcript. Specifically, insertion requires the target content to appear after the specified anchor when applicable; deletion requires the target content to be absent; and substitution requires the replacement content to appear while the original content is absent.

For content editing, SpeechEditBench defines Joint Success as Target Success, which corresponds to ES in our learning dynamics analysis. Given $N$ evaluation samples, the metrics are computed as
\begin{equation}
\mathrm{EM} =
\frac{1}{N}\sum_{i=1}^{N}
\mathbb{I}\left[
\mathcal{N}(y_i)=\mathcal{N}(\hat{y}_i)
\right],
\end{equation}
\begin{equation}
\mathrm{ES} = \mathrm{ACC} =
\frac{1}{N}\sum_{i=1}^{N}t_i,
\end{equation}
where $y_i$ and $\hat{y}_i$ denote the target and predicted transcripts, $\mathcal{N}(\cdot)$ denotes text normalization, and $t_i$ indicates whether the requested edit succeeds.

\paragraph{Paralinguistic Editing.}
SpeechEditBench evaluates the insertion and deletion of four nonverbal vocalizations: breath, laugh, cough, and sigh. Since the original Gemini 2.5 Pro judge is no longer available, we use Gemini 3.1 Pro Preview (\texttt{gemini-3.1-pro-preview}) as the audio judge, following the original scoring protocol. Each event is assigned a score from 0 to 3. Insertion is considered successful when the target event score is at least 2, while deletion succeeds when the score is at most 1.

In addition, linguistic content preservation is verified using ASR-based WER or CER, with a threshold of 10\%. The reported ACC corresponds to Joint Success, which requires both the target edit and linguistic content preservation to succeed:
\begin{equation}
\mathrm{ACC} =
\frac{1}{N}\sum_{i=1}^{N}t_i p_i,
\end{equation}
where $t_i$ and $p_i$ indicate target success and content preservation success, respectively.

\subsection{Ming-Freeform-Audio-Edit}

\paragraph{Editing Accuracy (ACC).}
ACC measures the proportion of successfully executed editing operations. The evaluator identifies the target edit region from the editing instruction or the differences between the source and target transcripts. For insertion and substitution, success requires the expected content to appear at the designated position. For deletion, success requires the original content at the target position to be removed or changed. ACC is computed as the proportion of successful edits across all evaluated samples.

\paragraph{WER and No-edit WER.}
WER measures the transcription error rate against the expected edited transcript. No-edit WER evaluates the preservation of linguistic content outside the requested edit. Following the official implementation, the evaluator constructs reference and hypothesis transcripts by handling the edited spans according to the operation, then computes the error rate over the resulting text sequences.

\paragraph{Speaker Similarity (SIM).}
SIM measures speaker identity preservation between the source and edited speech. Following the official benchmark, cosine similarity is computed between WavLM-Large\citep{chen2022wavlm} speaker embeddings extracted from the two utterances.

\section{Effect of Re-transcription on Speech Generation}
\label{app:retranscription}
To evaluate the effect of re-transcription on speech generation, we train two TTS models with identical architectures and training configurations from scratch on Emilia \citep{he2024emilia}, using either the original transcripts or those generated by Qwen3-ASR-1.7B \citep{shi2026qwen3asr}. Both models are trained for 1.1M updates, with the training transcripts being the only difference. We evaluate them on Seed-TTS test-zh and test-en \citep{anastassiou2024seedtts} and LibriSpeech-PC test-clean \citep{meister2023librispeechpc}, using word error rate (WER), speaker similarity (SIM), and UTMOS \citep{saeki2022utmos}.

As shown in Table~\ref{tab:retranscription}, re-transcription reduces WER by 3.5\%, 8.2\%, and 0.9\% relative on the three test sets, respectively, while SIM decreases by at most 0.007 and UTMOS by at most 0.10. These results suggest that re-transcription can improve the content accuracy of speech generation with only minor reductions in speaker similarity and predicted naturalness.

\begin{table}[t]
    \centering
    \small
    \setlength{\tabcolsep}{3pt}
    \begin{tabular}{@{}llccc@{}}
        \toprule
        Benchmark & Transcripts & WER (\%)$\downarrow$ & SIM$\uparrow$ & UTMOS$\uparrow$ \\
        \midrule
        \multirow{2}{*}{Seed-TTS zh}
        & Original & 1.462 & \textbf{0.756} & \textbf{3.01} \\
        & Qwen3-ASR & \textbf{1.411} & 0.751 & 2.91 \\
        \midrule
        \multirow{2}{*}{Seed-TTS en}
        & Original & 1.600 & \textbf{0.669} & \textbf{3.76} \\
        & Qwen3-ASR & \textbf{1.469} & 0.662 & 3.66 \\
        \midrule
        \multirow{2}{*}{LibriSpeech-PC}
        & Original & 2.146 & \textbf{0.664} & \textbf{3.93} \\
        & Qwen3-ASR & \textbf{2.126} & 0.657 & 3.88 \\
        \bottomrule
    \end{tabular}
    \caption{Effect of re-transcription on TTS performance. Both models are trained from scratch for 1.1M updates on Emilia, differing only in the training transcripts. Qwen3-ASR denotes Qwen3-ASR-1.7B.}
    \label{tab:retranscription}
\end{table}

\section{Additional Dataset Statistics}
\label{app:data-stats}

\begin{figure}[t]
    \centering
    \includegraphics[width=\linewidth]{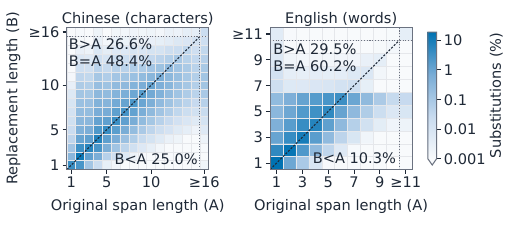}
    \caption{Joint distributions of original span length ($A$) and replacement span length ($B$) for substitution editing in Chinese and English. The color scale indicates the percentage of samples on a logarithmic scale.}
    \label{fig:substitution_matrix}
\end{figure}

Figure~\ref{fig:substitution_matrix} presents the joint distribution of original and replacement span lengths for substitution editing. Equal-length substitutions account for 48.4\% of Chinese and 60.2\% of English samples. Length-increasing substitutions account for 26.6\% and 29.5\%, respectively, while length-decreasing substitutions account for 25.0\% and 10.3\%. These statistics show that \dataset{} covers both length-preserving and length-changing substitutions, with a higher proportion of length-decreasing substitutions in Chinese.

\section{Model Configuration for Learning Dynamics}
\label{app:probe-config}

For the learning dynamics experiments, we adopt a smaller AuK-based model with approximately 278M trainable backbone parameters. The model retains the original architecture while reducing the Transformer depth and hidden dimension. Table~\ref{tab:probe-config} summarizes the architecture and training configurations.

\begin{table}[t]
\centering
\footnotesize
\setlength{\tabcolsep}{4pt}
\renewcommand{\arraystretch}{1.08}
\begin{tabular}{@{}lcc@{}}
\toprule
Configuration & AuK & Ours \\
\midrule
\multicolumn{3}{@{}l}{\textit{Transformer Backbone}} \\
MMDiT blocks & 10 & 4 \\
DiT blocks & 20 & 8 \\
Hidden dimension & 1536 & 1024 \\
Attention heads & 24 & 16 \\
FFN dimension & 3072 & 2048 \\
Backbone parameters & $\sim$1.5B & 278.31M \\
\midrule
\multicolumn{3}{@{}l}{\textit{Shared Configuration}} \\
Attention head dimension & \multicolumn{2}{c}{64} \\
FFN type & \multicolumn{2}{c}{SwiGLU} \\
Semantic encoder family & \multicolumn{2}{c}{Qwen2.5-Omni} \\
Semantic encoder & \multicolumn{2}{c}{Frozen} \\
Audio VAE & \multicolumn{2}{c}{AuK-VAE (frozen)} \\
VAE latent dimension & \multicolumn{2}{c}{64} \\
Latent frame rate & \multicolumn{2}{c}{50 Hz} \\
Audio sample rate & \multicolumn{2}{c}{24 kHz} \\
Time sampling & \multicolumn{2}{c}{Logistic-normal} \\
$P_{\mathrm{mean}}$ & \multicolumn{2}{c}{$-0.8$} \\
$P_{\mathrm{std}}$ & \multicolumn{2}{c}{$0.8$} \\
\midrule
\multicolumn{3}{@{}l}{\textit{Training Configuration}} \\
Semantic encoder & -- & Qwen2.5-Omni-3B \\
Optimizer & -- & AdamW (fused) \\
Learning rate & -- & $1\times10^{-4}$ \\
Adam $\beta_1, \beta_2$ & -- & $(0.9, 0.95)$ \\
LR scheduler & -- & Warmup + constant \\
Warmup steps & -- & 2,500 \\
Gradient clipping & -- & 1.0 \\
Gradient accumulation & -- & 1 \\
Number of GPUs & -- & 16 \\
Batch size per GPU & -- & 18,000 frames \\
Max. samples per GPU & -- & 16 \\
EMA decay & -- & 0.9999 \\
EMA update interval & -- & 10 steps \\
Training steps & -- & 200K (default) \\
\bottomrule
\end{tabular}
\caption{Model architecture and training configurations for the learning dynamics experiments. Shared architectural settings are merged. Training hyperparameters are reported for our smaller model only.}
\label{tab:probe-config}
\end{table}

\section{Detailed Convergence Curves of Editing Tasks}
\label{app:task-convergence}

\begin{figure*}[t]
    \centering
    \includegraphics[width=\textwidth]{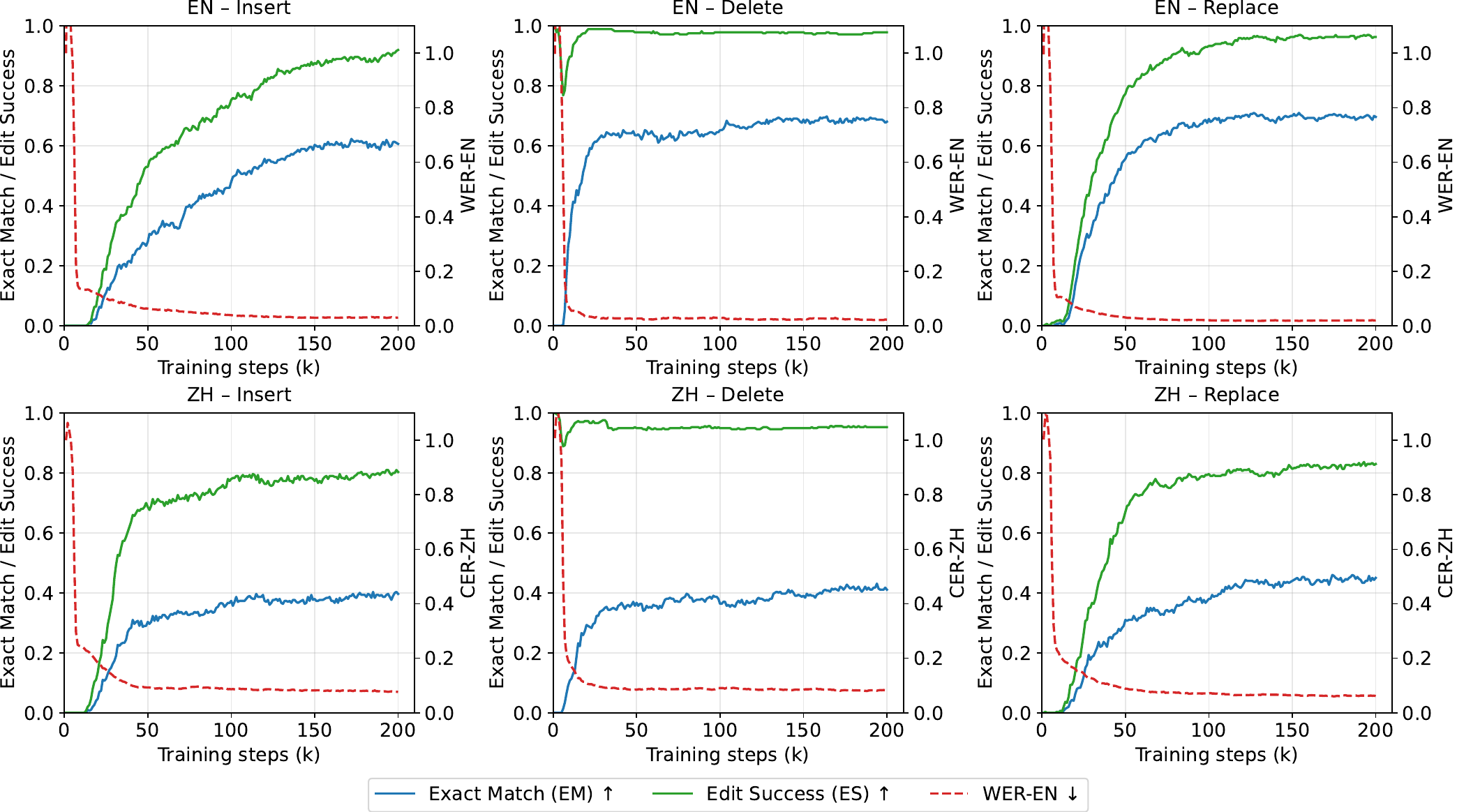}
    \caption{Detailed convergence curves of individual editing tasks in English (top) and Chinese (bottom). The three columns correspond to insertion, deletion, and substitution. Each panel reports Exact Match (EM), Edit Success (ES), and ASR error rate (WER for English and CER for Chinese) over 200K training steps.}
    \label{fig:task_convergence_detailed}
\end{figure*}

Figure~\ref{fig:task_convergence_detailed} presents the detailed convergence curves of insertion, deletion, and substitution in both English and Chinese. Both languages exhibit similar learning behaviors, with deletion converging the fastest, followed by substitution and insertion. However, the absolute performance differs between the two languages. Notably, the ES of deletion saturates much earlier than EM, indicating that operation-level success does not necessarily imply accurate reconstruction of the target transcript.

\section{Detailed Analysis of TTS Initialization}
\label{app:tts_initialization}

\begin{figure*}[t]
    \centering
    \includegraphics[width=\textwidth]{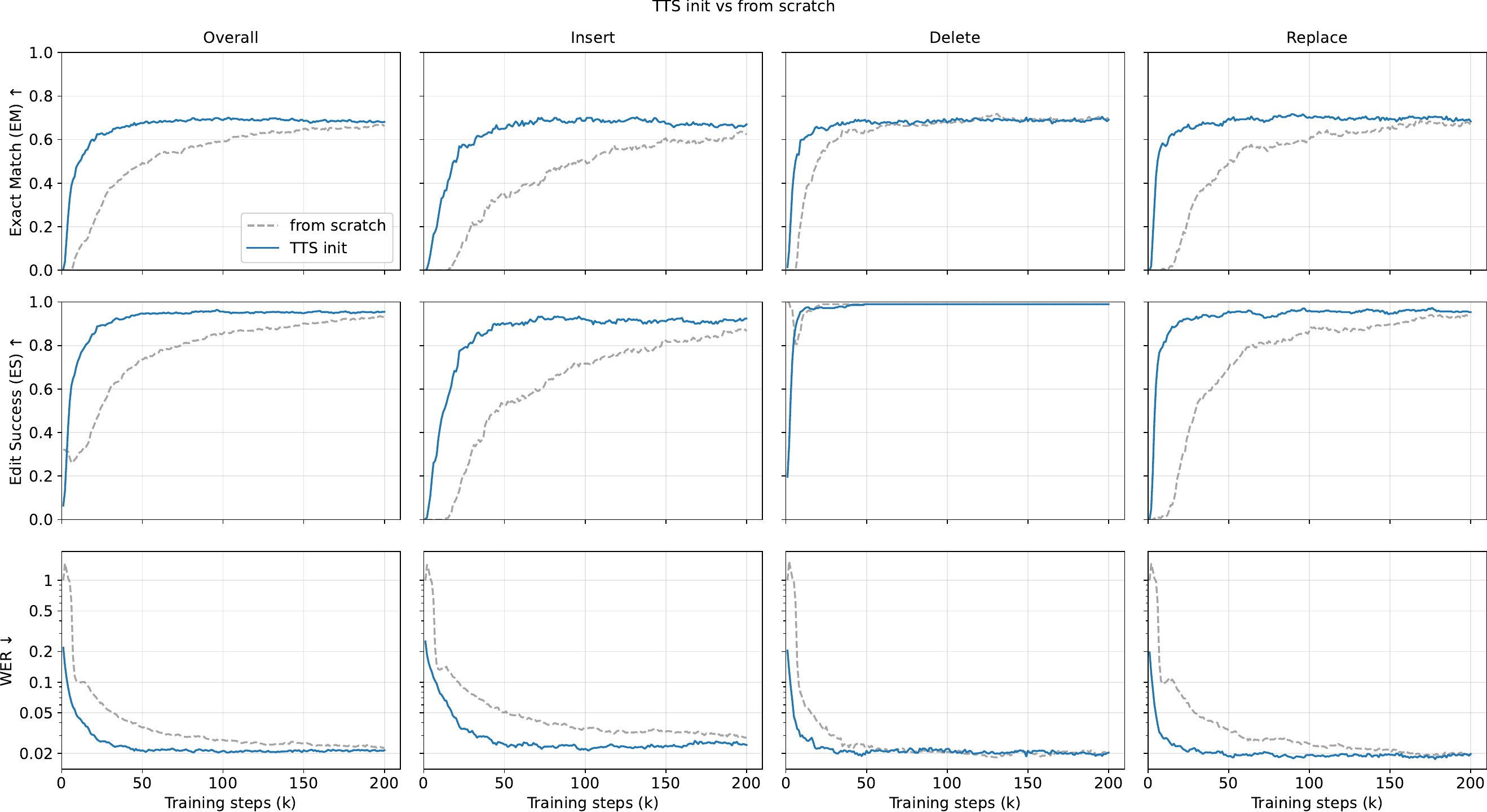}
    \caption{Detailed convergence curves of speech editing models with and without TTS initialization. Columns correspond to overall performance, insertion, deletion, and substitution. Rows report Exact Match (EM), Edit Success (ES), and Word Error Rate (WER), respectively.}
    \label{fig:tts_initialization_detailed}
\end{figure*}

Figure~\ref{fig:tts_initialization_detailed} presents the detailed convergence curves of editing models trained from scratch and initialized from a pretrained TTS checkpoint. TTS initialization accelerates convergence across all three editing operations, with particularly pronounced improvements in insertion and substitution. It also leads to a faster reduction in WER during early training. As training progresses, the performance gap gradually narrows, suggesting that TTS pretraining primarily improves learning efficiency rather than the final editing performance.

\section{Detailed Analysis of Data Mixture Ratios}
\label{app:mixture-ratios}

\begin{figure*}[t]
    \centering
    \includegraphics[width=\textwidth]{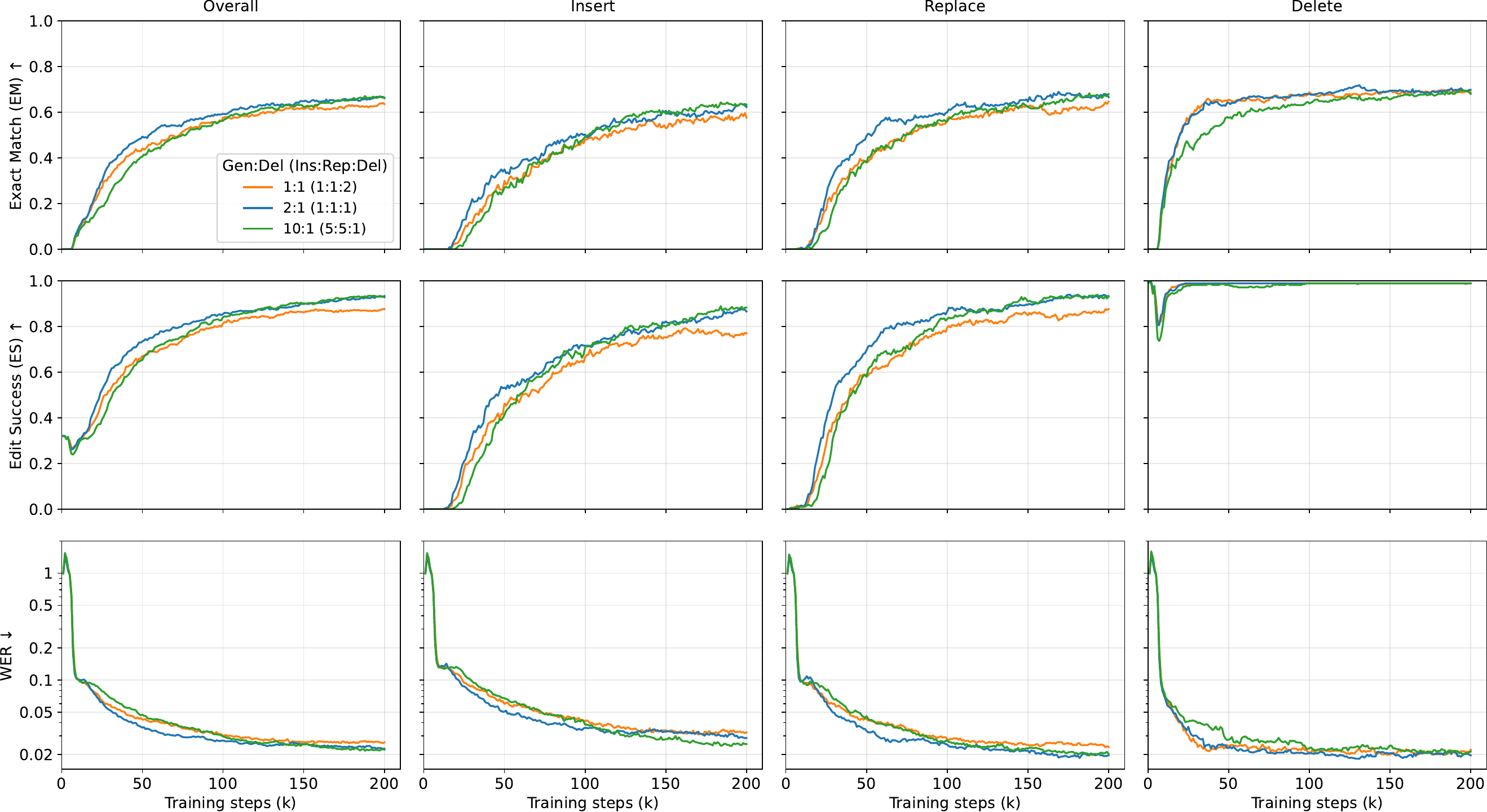}
    \caption{Detailed learning curves under different data mixture ratios. Columns correspond to overall performance, insertion, substitution, and deletion. Rows report Exact Match (EM), Edit Success (ES), and Word Error Rate (WER), respectively. The three generation-intensive-to-deletion ratios are 1:1 (1:1:2), 2:1 (1:1:1), and 10:1 (5:5:1), with the ratios in parentheses denoting insertion, substitution, and deletion.}
    \label{fig:mixture_ratios_detailed}
\end{figure*}

Figure~\ref{fig:mixture_ratios_detailed} presents the detailed learning curves under three generation-intensive-to-deletion ratios: 1:1 (1:1:2), 2:1 (1:1:1), and 10:1 (5:5:1), where the ratios in parentheses correspond to insertion, substitution, and deletion, respectively. The 2:1 mixture achieves the best overall trade-off across EM, ES, and WER. Increasing the proportion of deletion reduces insertion and substitution performance, while substantially reducing deletion slows early convergence without improving final performance. Notably, deletion ES remains close to saturation across all configurations, making EM more informative for comparing deletion performance. These results further demonstrate that balancing task proportions is important for effective joint training.

\section{Detailed Analysis of Curriculum Learning}
\label{app:curriculum-learning}

\begin{figure*}[t]
    \centering
    \includegraphics[width=\textwidth]{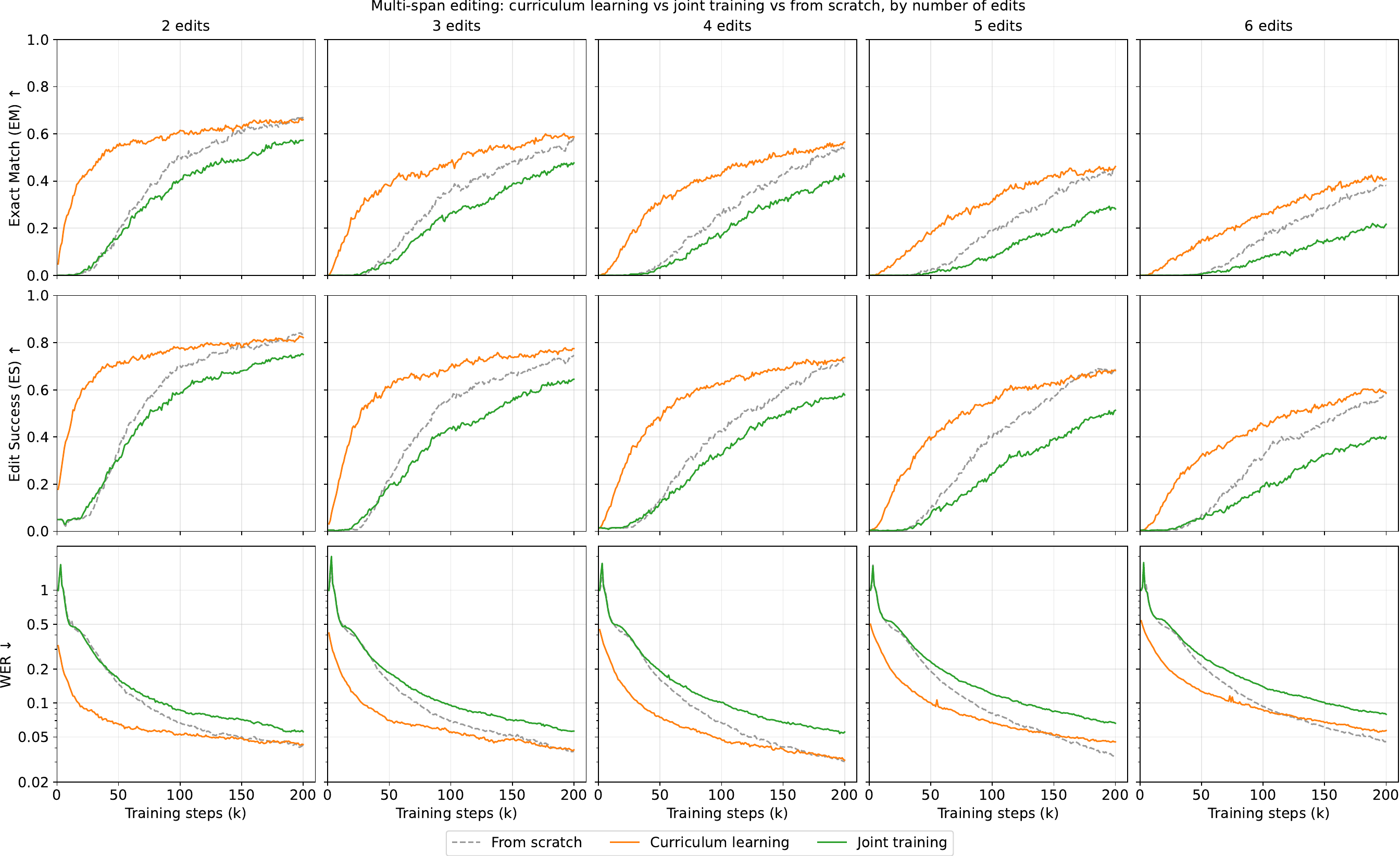}
    \caption{Detailed convergence curves of multi-span editing across different numbers of edits. The five columns correspond to 2, 3, 4, 5, and 6 edits, respectively. The top and bottom rows show Exact Match (EM) and Edit Success (ES).}
    \label{fig:curriculum_learning_detailed}
\end{figure*}

Figure~\ref{fig:curriculum_learning_detailed} presents the convergence curves for multi-span editing with 2 to 6 editing operations. Across all three training strategies, both EM and ES generally decrease as the number of edits increases, reflecting the growing difficulty of correctly executing multiple operations within a single instruction. Nevertheless, curriculum learning consistently accelerates convergence across different edit counts, demonstrating that previously learned single-span editing capabilities can facilitate the learning of more complex editing instructions.

\end{document}